\documentclass[5p,sort&compress,lefttitle]{elsarticle}
\usepackage{longtable,dcolumn}

\usepackage[utf8]{inputenc}
\usepackage[T1]{fontenc}
\usepackage{graphics,graphicx}
\usepackage{epstopdf}
\usepackage{float}
\usepackage{natbib}
\usepackage{subfigure}
\usepackage[colorlinks=true,linkcolor=blue,citecolor=blue,urlcolor=blue,pdfencoding=auto,psdextra]{hyperref}
\usepackage[figurename=Fig.,labelfont=bf]{caption}
\usepackage{pifont}
\usepackage[version=4]{mhchem}
\usepackage{color}

\usepackage{array}
\usepackage{siunitx}

\usepackage{xcolor}
\usepackage{listings}
\usepackage{bera}
\usepackage{multirow}

\DeclareCaptionFormat{listing}{\textbf{#1}#2#3}
\definecolor{numb}{rgb}{0.7,0.3,0.6}
\definecolor{keyw}{rgb}{0.3,0.7,0.3}
\definecolor{valu}{rgb}{0.2,0.5,0.8}
\newcommand\pythonstyle{\lstset{
    language=Python,
    tabsize             = 4,
    backgroundcolor     = \color{white},   
    keywordstyle        = \color{keyw}\bfseries,    
    numberstyle         = \tiny\color{numb},
    stringstyle         = \color{valu},
    basicstyle          = \ttfamily \scriptsize,
    showstringspaces    = false,
    keywords            = {false,true,null},
    alsoletter          = 0123456789.,
    string              = [s]{"}{"},
    breakatwhitespace   = false,         
    breaklines          = false,              
    keepspaces          = true,                 
    numbers             = none,                    
    numbersep           = 5pt,                  
    showspaces          = false,                
    showstringspaces    = false,
    escapeinside        = {(*@}{@*)},
    showtabs            = false,                  
    tabsize             = 2,
    keywordstyle        = \color{keyw},
    moredelim           = [is][\color{keyw}]{\%\%}{\%\%},
    literate            =
    *{0}{{{\color{numb}0}}}{1}
     {1}{{{\color{numb}1}}}{1}
     {2}{{{\color{numb}2}}}{1}
     {3}{{{\color{numb}3}}}{1}
     {4}{{{\color{numb}4}}}{1}
     {5}{{{\color{numb}5}}}{1}
     {6}{{{\color{numb}6}}}{1}
     {7}{{{\color{numb}7}}}{1}
     {8}{{{\color{numb}8}}}{1}
     {9}{{{\color{numb}9}}}{1}
     {"version"}{{{\color{keyw}"version"}}}{9}
     {"inchikey"}{{{\color{keyw}"inchikey"}}}{10}
     {"iso\_slug"}{{{\color{keyw}"iso\_slug"}}}{10}
     {"iso\_formula"}{{{\color{keyw}"iso\_formula"}}}{13}
     {"dataset\_name"}{{{\color{keyw}"dataset\_name"}}}{14}   
     {"name"}{{{\color{keyw}"name"}}}{6}
     {"desc"}{{{\color{keyw}"desc"}}}{6}
     {"ffmt"}{{{\color{keyw}"ffmt"}}}{6}
     {"cfmt"}{{{\color{keyw}"cfmt"}}}{6}      
     {"code"}{{{\color{keyw}"code"}}}{6}
     {"num\_lines"}{{{\color{keyw}"num\_lines"}}}{11}
     {"quantum\_numbers"}{{{\color{keyw}"quantum\_numbers"}}}{17} 
     {"Ka'"}{{{\color{valu}{"Ka'"}}}}{5}
     {"Kc'"}{{{\color{valu}{"Kc'"}}}}{5}
     {"v1'"}{{{\color{valu}{"v1'"}}}}{5}
     {"v2'"}{{{\color{valu}{"v2'"}}}}{5}
     {"v3'"}{{{\color{valu}{"v3'"}}}}{5}
     {"Ka\\""}{{{\color{valu}{"Ka\textbackslash""}}}}{6}
     {"Kc\\""}{{{\color{valu}{"Kc\textbackslash""}}}}{6}
     {"v1\\""}{{{\color{valu}{"v1\textbackslash""}}}}{6}
     {"v2\\""}{{{\color{valu}{"v2\textbackslash""}}}}{6}
     {"v3\\""}{{{\color{valu}{"v3\textbackslash""}}}}{6}
     {"J'"}{{{\color{valu}{"J\textquotesingle"}}}}{4}
     {"Ka'"}{{{\color{valu}{"Ka\textquotesingle"}}}}{5}
     {"Kc'"}{{{\color{valu}{"Kc\textquotesingle"}}}}{5}
     {"v1'"}{{{\color{valu}{"v1\textquotesingle"}}}}{5}
     {"v2'"}{{{\color{valu}{"v2\textquotesingle"}}}}{5}
     {"v3'"}{{{\color{valu}{"v3\textquotesingle"}}}}{5}  
}
}
\lstnewenvironment{python}[1][]{
\pythonstyle \lstset{#1} }{}

\graphicspath{{figure/}}

\newcommand{\cm}{cm$^{-1}$}
\newcommand{\Marvel}{{\sc MARVEL}}

\newcommand{\tick}{\ding{52}}
\newcommand{\nmol}{91}
\newcommand{\niso}{224}
\newcommand{\HDOname}{TBD}
\newcommand{\ai}{\textit{ab initio}}

\journal{Journal of Quantitative Spectroscopy \& Radiative Transfer}
\begin{document}

\begin{frontmatter}

\title{The 2024 release of the ExoMol database: molecular line lists for exoplanet and other hot atmospheres}
\author[UCL]{Jonathan Tennyson\corref{cor}}
\ead{j.tennyson@ucl.ac.uk}
\author[UCL]{Sergei N. Yurchenko}
\author[UCL]{Jingxin Zhang}
\author[UCL]{Charles A. Bowesman}
\author[UCL]{Ryan P. Brady}
\author[JB]{Jeanna Buldyreva}
\author[Bris]{Katy L. Chubb}
\author[RRG]{Robert R. Gamache}
\author[Bris,Sussex]{Maire N. Gorman}
\author[UCL]{Elizabeth R. Guest}
\author[IAEA]{Christian Hill}
\author[UCL]{Kyriaki Kefala}
\author[UCL,OXF,UWC]{A. E. Lynas-Gray} 
\author[UCL]{Thomas M. Mellor}
\author[UNSW]{Laura K. McKemmish}
\author[UCL]{Georgi B. Mitev}
\author[NN]{Irina I. Mizus}
\author[UCL]{Alec Owens}
\author[UCL]{Zhijian Peng}
\author[UCL]{Armando N. Perri}
\author[UCL,Peruga]{Marco Pezzella}
\author[UCL,NN]{Oleg L. Polyansky}
\author[UCL]{Qianwei Qu}
\author[UCL]{Mikhail Semenov}
\author[UCL]{Oleksiy Smola} 
\author[UCL]{Andrei Solokov}
\author[UCL]{Wilfrid Somogyi}
\author[UCL]{Apoorva Upadhyay}
\author[UCL]{Samuel O.M. Wright}
\author[NN]{Nikolai F. Zobov}

\cortext[cor]{Corresponding author.}

\address[UCL]{Department of Physics and Astronomy, University College London, Gower Street, London WC1E 6BT, UK}
\address[Bris]{School of Physics, HH Wills Physics Laboratory, Tyndall Avenue, Bristol, BS8 1TL, UK}
\address[OXF]{Department of Physics, University of Oxford, Keble Road, Oxford OX1 3RH, UK}
\address[UWC]{Department of Physics and Astronomy, University of the Western Cape, Private Bag X17, Bellville 7535, South Africa}
\address[IAEA]{Nuclear Data Section, International Atomic Energy Agency, Vienna A-1400, Austria}
\address[UNSW]{School of Chemistry, University of New South Wales, 2052 Sydney, Australia}
\address[JB]{Institut UTINAM UMR CNRS 6213, Universit\'{e} de Franche-Comt\'{e}, 16 Route de Gray, 25030 Besan\c{c}on cedex, France}
\address[RRG]{Department of Environmental, Earth, and Atmospheric Sciences University of Massachusetts Lowell, Lowell, MA 01854 USA}
\address[Perugia]{Dipartimento di Fisica e Geologia,  Universit\'{a} di Perugia, Via Pascoli SNC,06123 Perugia, Italy}
\address[Sussex]{School of Mathematical and Physical Sciences, University of Sussex, Brighton, BN1 9QH, UK}
\address[NN]{Institute of Applied Physics, Russian Academy of Sciences, 46 Ulyanov Street, Nizhny Novgorod, 603950, Russia}
\date{\today}

\begin{abstract}

The ExoMol database (www.exomol.com) provides molecular data for spectroscopic studies of hot atmospheres. These data are widely used to model atmospheres of exoplanets,  cool stars and other astronomical objects, as well as a variety of terrestrial applications. The 2024 data release reports the current status of the database which contains recommended line lists for \nmol\ molecules and \niso\ isotopologues giving a total of almost 10$^{12}$ individual transitions.  New features of the database include extensive "MARVELization" of line lists to allow them to be used for high resolutions studies, extension of several line lists to ultraviolet wavelengths, provision of photodissociation cross sections and extended provision of broadening parameters. Some of the in-house data specifications have been rewritten in JSON and moved to conformity with other international standards. Data products, including specific heats, a database of lifetimes for plasma studies, and the ExoMolHR web app which allows  exclusively high resolution data to be extracted, are discussed.

\end{abstract}

\end{frontmatter}
\newpage

\onecolumn
\section{Introduction}

The ExoMol project was founded in 2011 to provide molecular line lists for exoplanet and other atmospheres \cite{jt528} with a particular emphasis on providing results for
elevated temperatures which are not adequately covered by databases such as HITRAN \cite{jt857}
or GEISA \cite{jt632}. ExoMol has followed the practice of providing data
release papers every four years with the first two in 2016 \cite{jt631} and 2020 \cite{jt810};
the present paper is the third such release. A short review of the first decade of
activity by the ExoMol project has been given by the project founders Tennyson and Yurchenko \cite{ExoMol10}.

Since the launch of the project, ExoMol has created line lists for 67 molecules (approaching 150 isoptopologues) details of which are given below. Also discussed
below are line lists created
by other teams, which are included in the database to ensure coverage of key molecular species as far as possible. A major goal  of the original ExoMol project was to provide
line lists which were as complete as possible as these have been shown to be
essential for correctly reproducing broad band spectra \cite{jt572} and also opacities. Since that time, high resolution Doppler-shift cross-correlation spectroscopy has been pioneered as a very successful tool for detecting atoms and molecules in exoplanet atmospheres
\cite{04Snellen.exo,14Snellen.exo,21BrBi.exo}. Key line lists constructed with an emphasis on completeness lacked the necessary accuracy for these high resolution studies
\cite{15HoDeSn.exo,22deKeSn.VO}, see also Yurchenko {\it et al.} \cite{jt940}, thus addressing the needs of high resolution observation led to a significant extension in the scope of the ExoMol project.


The tension between completeness and accuracy in providing hot
line lists for astronomical modelling is well-known \cite{jt716}.
ExoMol has developed techniques based on the use of the MARVEL (measured active vibration-rotation energy levels) algorithm \cite{jt412} to
provide empirical energy levels from  high resolution laboratory spectra. High accuracy energy levels plus extrapolation from them \cite{jt948} can be used to provide accurate transitions while retaining the completeness of the computed line list. These procedures are discussed in Sect.~\ref{ss:MARVEL}. 

The ExoMol project has continued to provide line lists for new species and has updated and/or extended existing line lists for a number of other molecules. The database has  also developed in other directions: notably there has been increasing provision for photoabsorption at shorter wavelengths. Extending the database to ultraviolet wavelengths has led to the need to consider not only line absorption at these wavelengths but also continuum absorption, line broadening due to predissociation and photodissociation rates. As a result we have had to both develop new methods for treating photodissociation \cite{jt840}, predissociation \cite{jtpred,jt922,jt933} and continuum absorption \cite{jt922}, and to extend the ExoMol data model to allow for these processes \cite{jt898}. 

Since the ExoMol 2020 release, ExoMol line lists have been used extensively in the modelling of gas giant atmospheres, such as sub-Neptunes \cite{20HoFoLi.exo,20PeMaxx.exo,21TsInLi.exo,21BlChBe.exo,21ChBlBe.exo,22GaMiCh.exo,22BaWoTe.exo,22ZaHeDr.exo,22CoMaxx.exo,22GuKiMo.exo,23LoZhHu.exo,23GaPiSt.exo,23KeZhBe.exo,23MiMaDi.exo,24HoMaxx.exo,24DaBrGa.exo}, Neptune-like exoplanets \cite{20GuKiFi.exo,20GuGrWr.exo,23AtAwJi.exo,23EdChTs1.exo,24RaCoTa.exo}, so-called Hycean planets \cite{21MaPiCo.exo,23MaSaCo.exo,23KeLeMa.exo}, Saturn-like exoplanets \cite{20SpSiWa.exo,20AnEdCh.exo,20ChEdAl.exo,20CoKrWe.exo,22MaBrCh.exo,23AhAlBa.exo,22NiSiSp.exo,23TsLePo.exo,23SaHeCh.exo,23RaWeEs.exo,23AlWaAl.exo,23FoMaRa.exo,23TaRaWe.exo,23BoLaPe.exo}, young Jupiters \cite{23WhGlCh.dwarfs}, cool Jupiters \cite{21XuDiMa.exo}, warm Jupiters \cite{22GaMiCh.exo,22GuGiCa.exo,23ZhMoHa.exo,22CaGiGu.exo,23BeWeSc.exo,23KaMaLi.exo}, hot Jupiters \cite{20ZhChKe.exo,20PiMaMc.TiO,20YiChEd.exp,20LeCzDe.exo,20LaBuxx1.exo,20ChCaPa.exo,20WiKoBa.exo,21GiBrGa.exo,21CaPaSt.exo,21SeSnMo.exo,21ShWeMa.exo,21BrVaCh.exo,21ChAlEd.exo,21RaMaBa.exo,21IsKaNu.exo,21CuKeCo.exo,21LePaHa.exo,21BaDeCa.exo,21RuKoBa.exo,21SeMaCa.exo,21LiBrBe.exo,21WaRuMo.exo,22WeMaxx.exo,22ChMixx.exo,22SaTsMo.exo,22ChEdAl.exo,22ScCaDe1.exo,22ScCaDe2.exo,22AlChVe.exo,22KaBeLi.exo,22WeGaBr.exo,22DeHeSc.exo,22BlKaPi.exo,22PlLePa.exo,22ZaHeDr.exo,22NiMaxx.exo,23SpNiCo.exo,23BaPeBe.exo,23RaBuMe.exo,23WaChTi.exo,23CuFoKo.exo,23BeKnMe.exo,23ChStHe.exo,23RuSiMu.exo,23LoMiKu.exo,23BaBaHa.exo,23Crossfield.exo,23CoMaGa.exo,23FeRaWe.exo,23GrLeWa.exo,23GadeSn.exo,23ZaPeMc.exo,23WeMcLi.exo,23TaPaxx.exo,23ChBuSu.exo,23HaCuQu.exo,23YaIrBa.exo,23ChMaHo.exo,23AhStMa.exo,23LeNoYa.exo,24EdChxx.exo,24SmLiBe.exo,24ScMoLo.exo,24BeCrCu.exo,24PoFeLe.exo,24YiChAl.exo,24LaStSn.exo} and ultra-hot Jupiters \cite{20WiGiNi.exo,20GaJexx.exo,21TaZaAl.exo,21MaKoSt.exo,21GhIyLi.exo,21StPaCa.exo,21KaBeLi.exo,21RoDrHe.exo,21SeNuMo.exo,21KiRaMa.exo,21ChEdxx.exo,22LePrKi.exo,22BiHoKi.exo,22MiSiBa.exo,22SaKeMa.exo,22BeCaMe.exo,22Changeat.exo,22deKeSn.VO,22HoMaxx.exo,23FiScSa.exo,23YaChWa.exo,23LoZhWr.exo,23WaPaLi.exo,23CoBeCh.exo,23JoWaAs.exo,23PeJoAs.exo,23KiHoGr.exo,23PeBeAl.exo,23ShWaZh.exo,23RideJa.exo,23GaKeZh.exo,23BrEmLi.exo,23PrHoPe.exo,23vaBiLo.exo,24ChSkCh.exo,24FiXuXi.exo,24MaGiNu.exo}.
They have also been employed in characterising the atmospheres of terrestrial planets \cite{20EdChMo.exo,21PiMaMa.exo,21MuMoEd.exo,22GrMoCh.exo,22AlKoAn.exo,22BoHaSo.exo,23ClRiRu.exo,23DiMeCh.exo,23ZhMaWa.exo,23PhWaEd.exo,23MoStSi.exo}, particularly hot rocky exoplanets \cite{22ZhMaWa.exo,22ZiVaMi.exo,22WhMaIh.exo,22CrMaHi.exo,22ZiZiKr.exo,23JaWoHe.exo,23PiGaBr.exo,23Heng.exo,23RiNuFl.exo}, those containing volatiles \cite{23ZiMiBu.exo} and those undergoing photoevaporation \cite{23BoOwSc.exo,24SeFeLa.exo}.
Within our Solar System, they have been used to study the Venusian atmosphere \cite{22GrRiRi.SO2,22DiMaRi.SO2}, volcanic plumes on Io \cite{21KoSuTs.SO2} and comets \cite{21CaCrMuCa}.
In all kinds of planetary atmospheres they have been utilised in the modelling of clouds and hazes \cite{20HoFoLi.exo,20LaBuxx1.exo,20LaBuxx2.exo,21WeMaxx.exo,21LaArxx.exo,21MuBaMa.exo,21ChBlBe.exo,22LeTaLe.dwarfs,23SaHeCh.exo,23JiChPa.exo,23GrLeWa.exo,23SpNiCo.exo,23RaBuMe.exo,23LaBuxx.dwarfs,23MaItAl.exo} and photochemistry \citep{22BaKoVe.exo}.
ExoMol line lists have been used to test the detection capabilities of current and future space based telescopes \cite{21EdStxx.exo,21MuAlBo.exo,22ChMeKr.exo,22GuSoGi.exo,22GaMiCh.exo,22EdTixx.exo,22BaChCh.exo,23EdChTs2.exo,23BoMuPa.exo,23PhWaEd.exo,24ChItAl.exo} and have frequently been incorporated in a variety of radiative transfer, forward-modelling, spectral synthesis and opacity codes \cite{TRIDENT,jt611,TauRex3,22AlChVe.exo,BART,BARTII,gCMCRT,21CuBlxx.exo,MALBEC,ARCiS,RAPOC,24WhCaAb.exo}.
They have also been employed to study the detectability of biosignatures \citep{21ZhSePe.exo} and prebiosignatures \citep{23ClRiRu.exo}.
In stars, ExoMol line lists have also aided in the study of molecular species in the atmospheres of main sequence dwarfs \cite{20IsAoKo.dwarfs,21CodeRe.stars,21MaTaMo.stars,22LyYuPa.NaH,23JoPaLy.stars,23CrDoFo.stars,24XuWaFi.stars}, giants \cite{20PaEvBa.stars,23GaArSl.stars,23HaVeWa.stars}, brown dwarfs \cite{20BrDuBo.dwarfs,20PiMaxx.dwarfs,21MeScBu.dwarfs,21voMaPi.dwarfs,21MuFoJe.dwarfs,21MaSaVi.dwarfs,21CuScKi.dwarfs,22LuKiBo.dwarfs,22TaMeHo.dwarfs,22LeTaLe.dwarfs,22XuWaRu.dwarfs,23LaBuxx.dwarfs,23RoMoLi.dwarfs,23BaMoPa.dwarfs,PICASO,23HoFoLi.dwarfs,23FrBoBo.dwarfs,23WhGlCh.dwarfs,23LeTaTs.dwarfs,24MuFoMo.exo,24BuBeLa.dwarfs,24LeRoBa.dwarf}, eruptive Young Stellar Objects \cite{jt928}, in stellar winds from Oxygen-rich \cite{20DaGoDe.stars,23CoDaDe.stars,23SaMaCh.stars} and Carbon-rich \cite{20PaMaGi.stars,23ErHoAr.stars} Asymptotic Giant Branch stars, in cool Carbon stars \cite{22KaKaTi.stars,23GaKaLa.stars}, in supergiants \cite{20DaGoDe.stars}, in stellar merger remnants \cite{21MoKaMa.stars,24StKaSc.stars} and modelling supernova ejecta \cite{23LiJeBa.stars}.
They have also been used to look at molecules in the interstellar medium (ISM) \citep{21NiRaCo.ISM,23NiRaCo.ISM} and to study the isotope exchange rates of molecules important in those environments \citep{21BoHiLi.H3+}.
Outside of astronomy, ExoMol line lists have been employed in laser absorption spectroscopy \cite{21LoWeSi.AlO,22DaMuAl.AlO,22MuAlBa.AlO,23GiRuDa.NO}, laser induced breakdown spectroscopy \cite{23AbLyGl.AlO,23RiHuYo.AlO},
laser fragmentation \cite{22BoGoZh.PO}, laser induced fluorescence \cite{22BoGoEv.PO}, plasma stoichiometry \cite{23MeNiEt.TiO}, fusion plasmas \cite{24PaBoBr.BeH}, the design of gas sensors \cite{24YeZaRo},
combustion \cite{21RuMcMa.AlO} and explosions \cite{23Fateev.CaOH}.
ExoMol line lists are often used as a benchmark in \textit{ab initio} calculations \cite{21QiBaLi.AlH,21daBaVe.NO,22ZaLeCo.NO,23ChStDe.CH3Cl,23BaQiLi.AlO,23VeLa.NO} and to compare line strengths \cite{21UlGrBe2.SiH4,21ShDoRo.H3+,21ZnGrWo.H3+,21FoRuSi.CH4,22LaVe.NO,23PeMaAr1.HNO3,23PeMaAr2.HNO3,23VaNaHo.SO2}.

The ExoMol database is complete enough to be used to generate other data products.
Examples include the creation of the Lifetimes DataBase (LiDB)  of vibronic state lifetimes, primarily for use in plasma modelling \cite{jt904}, see Sect.~\ref{ss:LiDB}; a database of NASA polynomial fits to
specific heats generated from ExoMol data \cite{jt899} and a database of high resolution transitions suitable for spectral assignment \cite{jtHR}, see Sect.~\ref{ss:HR}.
Importantly ExoMol line lists have been used to give molecular opacities in the ExoMolOP database \cite{jt801} discussed below; opacities based on ExoMol data have also been generated by several other groups \cite{14FrLuFo.OP,16AmMaBa.OP,17Min.OP,18GoMaSi.OP,18ViSmPr,jt782,jt819,21GhIyLi.exo,22CuHaBl.OP,22HiHaMi.OP,22MaArGi.OP,SPHINX,PICASO}.
The generation of opacities, see Sect.~\ref{ss:op}, requires treatment of line broadening, progress on this topic for the ExoMol project is discussed in Sect.~\ref{s:broad}. 
ExoMol is providing input to JWST (James Webb Space Telescope) through the MAESTRO (Molecules and Atoms in Exoplanet Science: Tools and Resources for Opacities) database, see \url{https://science.data.nasa.gov/opacities/} and the  exoplanet characterisation Ariel Space Mission \cite{jt946}, which is due to launch in 2029.

Of course ExoMol is not the only source of spectroscopic data for studies of atmospheres. HITRAN \cite{jt841} is a well established database \cite{21Rothman} which is designed for use at temperatures in the region of 296 K and contains molecules of importance in the Earth's atmosphere. HITEMP \cite{jt480} extends
HITRAN to higher temperatures, albeit currently for only eight molecules. Where appropriate the extension of HITRAN  or HITEMP is based on ExoMol data \cite{jt763}, and
ExoMol uses HITRAN line lists for the diatomic molecules HF, HCl, HBr and CO for which the HITRAN compilation is appropriate for use at higher temperatures.
The TheoRets database \cite{TheoReTS} contains very extensive computed line lists for nine polyatomic molecules with five or more atoms; not all of these line lists are suitable for high temperature studies. The NASA Ames group provides very extensive line lists for 6 molecules; 5 triatomics plus ammonia. Their CO$_2$ line lists, which include isotopologues, are designed for use at temperatures up to 3000 K \cite{23HuFrTa.CO2}. The SO$_2$ ExoMol line list
\cite{jt635} was produced in collaboration with the NASA Ames group. The MoLLIST \cite{MOLLIST} line lists due to Bernath and co-workers provides
empirical line lists designed for use at higher temperatures for 26 diatomics, ammonia and methane as well as cross-sections for eight polyatomic
molecules with six or more atoms. A number of the MoLLIST diatomic line lists are also available from the ExoMol database, see Sect.~\ref{ss:other}. 
The VALD \cite{VALD3} and Kurucz \cite{11Kurucz.db} databases, which are largely aimed at stellar atmospheres, contain data on some diatomic species.

As discussed in Sect.~\ref{s:photo}, with this release ExoMol has started providing temperature-dependent photodissociation cross sections. The main current provider
of photodissociation cross sections for astronomical studies is the Leiden VUV cross section database \cite{Leiden} whose data are largely designed  for use in studies of the (cold) interstellar medium. Thus far Leiden have not considered the temperature of the molecule in their photodissociation cross sections as the data are largely aimed for use at temperature of 300 K or below.

\section{Line lists}


Table~\ref{tab:exomoldata} summarizes the recommended ExoMol line lists present in the database. 
Note that a number of ExoMol line lists have been replaced by either extended and/or improved upon line lists for the same species; these older line lists are no longer recommended and are not listed in the table although they are still available via the ExoMol website.
The 2020 data release introduced the use of uncertainties in the energies given in the \texttt{.states} file to allow high resolution transitions to be identified. This feature has now been
introduced for all ExoMol-generated recommended line lists, although uncertainties are generally not provided for old line lists or those provided from other sources. At the same time, the previously optional lifetime column in the
\texttt{.states} file has been made mandatory and all \texttt{.states} files for recommended line lists include lifetimes; this feature is important for modelling line-broadening effects due to predissociation \cite{jt898}.

Below, specific notes for some individual line lists are given. Those line lists which were included in the 2020 release and have not been updated are not explicitly discussed, please refer to the 2020 manuscript or the cited original line list paper for further details. An overview of the temperature-dependent cross sections for earlier ExoMol line lists is already 
available \cite{jt731} and below we give similar plots for many of the newly computed line lists.
Before discussing individual line lists we discuss our strategy for improving the accuracy of calculated line positions.

\begin{table}[H]
{\centering
\footnotesize
\caption{Datasets created by the ExoMol project and included in the ExoMol database: recommended line lists only. Line lists denoted with a \tick\ are suitable for high resolution studies.}
\label{tab:exomoldata}
\begin{tabular}{llrrrrlcl}
\hline
Paper & Molecule & $N_{\rm iso}$ & $T_{\rm max}$ & $N_{\rm elec}$ & $N_{\rm lines}$$^a$ & DSName & & Reference\\
\hline
III & HCN/HNC & 1$^a$ & 4000 & 1 & 34~418~408 & Harris & \tick & \cite{jt570}\\
V & NaCl & 2 & 3000 & 1 & 702~271 & Barton & & \cite{jt583}\\
V & KCl & 4 & 3000 & 1 & 1~326~765 & Barton & & \cite{jt583}\\
VII & PH$_3$ & 1 & 1500 & 1 & 16~803~703~395 & SAlTY & & \cite{jt592}\\
VIII & H$_2$CO & 1 & 1500 & 1 & 12~688~112~669 & AYTY & \tick & \cite{jt597}\\
IX & AlO & 4 & 8000 & 3 & 4~945~580 & ATP & \tick & \cite{jt598}\\
X & NaH & 2 & 7000 & 2 & 79~898 & Rivlin & & \cite{jt605}\\
XI & HNO$_3$ & 1 & 500 & 1 & 6~722~136~109 & AlJS & & \cite{jt614}\\
XII & CS & 8 & 3000 & 1 & 548~312 & JnK & \tick & \cite{jt615}\\
XIII & CaO & 1 & 5000 & 5 & 21~279~299 & VBATHY & \tick & \cite{jt618}\\
XIV & SO$_2$ & 1 & 2000 & 1 & 1~300~000~000 & ExoAmes & \tick & \cite{jt635}\\
XV & H$_2$O$_2$ & 1 & 1250 & 1 & 20~000~000~000 & APTY & & \cite{jt638}\\
XVI & H$_2$S & 1 & 2000 & 1 & 115~530~373 & AYT2 & \tick & \cite{jt640}\\
XVII & SO$_3$ & 1 & 800 & 1 & 21~413~927~818 & UYT2 & \tick & \cite{jt641}\\
XIX & H$_2$$^{17,18}$O & 2 & 3000 & 1 & 519~461~789 & HotWat78 & \tick & \cite{jt665}\\
XX & H$_3^+$ & 1 & 3000 & 1 & 127~542~657 & MiZATeP & \tick & \cite{jt666}\\
XXII & SiH$_4$ & 1 & 1200 & 1 & 62~690~449~078 & OYT2 & & \cite{jt701}\\
XXIII & PO & 1 & 5000 & 1 & 2~096~289 & POPS & & \cite{jt703}\\
XXIII & PS & 1 & 5000 & 3 & 30~394~544 & POPS & & \cite{jt703}\\
XXIV & SiH & 4 & 5000 & 3 & 1~724~841 & SiGHTLY & & \cite{jt711}\\
XXV & SiS & 12 & 5000 & 1 & 91~715 & UCTY & & \cite{jt724}\\
XXVI & NS & 6 & 5000 & 1 & 3~479~067 & SNaSH & & \cite{jt725}\\
XXVII & C$_2$H$_4$ & 1 & 700 & 1 & 49~841~085~051 & MaYTY & & \cite{jt729}\\
XXIX & CH$_3$Cl & 2 & 1200 & 1 & 166~279~593~333 & OYT &  & \cite{jt733} \\
XXX & H$_2$$^{16}$O & 1 & 5000 & 1 & 5~745~071~340 & POKAZATEL & \tick & \cite{jt734}\\
XXXI & C$_2$ & 3 & 5000 & 8 & 6~080~920 & 8states & \tick & \cite{jt736}\\
XXXII & MgO & 5 & 5000 & 5 & 72~833~173 & LiTY & \tick & \cite{jt759}\\
XXXIII & TiO & 5 & 5000 & 13 & 58~983~952 & Toto & \tick & \cite{jt760}\\
XXXIV & PH & 1 & 4000 & 2 & 65~055 & LaTY & \tick & \cite{jt765}\\
XXXV & NH$_3$ & 1 & 1500 & 1 & 16~941~637~250 & CoYuTe & \tick & \cite{jt771}\\
XXXVI & SH & 2 & 3000 & 2 & 572~145 & GYT & \tick & \cite{jt776}\\
XXXVII & HCCH & 1 & 2000 & 1 & 4~347~381~911 & aCeTY & \tick & \cite{jt780}\\
XXXVIII & \ce{SiO2} & 1 & 3000 & 1 & 32~951~275~437 & OYT3 &  & \cite{jt797}\\
XXXIX & \ce{CO2} & 1 & 3000 & 1 & 7~996~570~390 & UCL-4000 & \tick & \cite{jt804}\\
XL & \ce{H3O+} & 1 & 1500 & 1 & 2~089~331~073 & eXeL & \tick & \cite{jt805}\\
XLI & \ce{KOH} & 1 & 3500 & 1 & 38~362~078~911 & OYT4 & & \cite{jt820}\\
XLI & \ce{NaOH} & 1 & 3500 & 1 & 49~663~923~092 & OYT5 & & \cite{jt820}\\
XLII & \ce{NO} & 1 & 3500 & 4 & 4~596~666 & XABC & \tick & \cite{jt831}\\
XLIII & \ce{NaO} & 1 & 2500 & 2 & 4~726~283 & NaOUCMe & \tick & \cite{jt854}\\
XLIV & \ce{SiO} & 1 & 3500 & 10 & 91~395~763 & SiOUVenIR & \tick & \cite{jt847}\\
XLV & \ce{CaH} & 1 & 5000 & 3 & 293~151 & XAB & \tick & \cite{jt858}\\
XLV & \ce{MgH} & 3 & 5000 & 3 & 88~575 & XAB & \tick & \cite{jt858}\\
XLVI & \ce{SiN} & 4 & 3000 & 6 & 43~646~806 & SiNful & \tick & \cite{jt858}\\
XLVII & \ce{CaOH} & 1 & 3000 & 3 & 24~215~753~701 & OYT6 & \tick & \cite{jt858}\\
XLVIII & \ce{H2CS} & 1 & 2000 & 1 & 43~56~116~660 & MOTY & \tick & \cite{jt886}\\
XLIX & \ce{AlCl} & 2 & 5000 & 4 & 4~722~048 & YNAT & \tick & \cite{jt887}\\
L & \ce{H3+} & 4 & 3000 & 1 & - & MiZo & \tick & \cite{jt890}\\
LI & \ce{LiOH} & 2 & 3000 & 1 & 331~274~717 & OYT7 & & \cite{jt905}\\
LII & \ce{CH+} & 2 & 5000 & 2 & 34~194 & PYT & & \cite{jt913}\\
LIII & \ce{YO} & 3 & 5000 & 6 & 60~678~140 & BRYTS & & \cite{jt921}\\
LIV & \ce{AlH} & 4 & 5000 & 2 & 36~152 & AloHa & \tick& \cite{jt922}\\
LV & \ce{VO} & 1 & 5400 & 15 & 58~904~173~243 & HyVO & \tick & \cite{jt923}\\
LVI & \ce{SO} & 1 & 5000 & 8 & 7~008~190 & SOLIS & \tick & \cite{jt924}\\
LVII & \ce{CH4} & 1 & 2000 & 1 & 50~395~644~806 & MM & \tick & \cite{jt926}\\
LVIII & \ce{OCS} & 1 & 2000 & 1 & 2~482~380~391 & OYT8 & \tick & \cite{jt943}\\
LIX & \ce{N2O} & 5 & 2000 & 1 & 1~360~351~722 & TYM & \tick & \cite{jt951}\\
LX & $^{15}$NH$_3$ & 1 & 1000 & 1 & 929~795~249 & CoYuTe-15 & \tick & \cite{jt952}\\
LXI & \ce{OH} & 1 & 5000 & 8 & 1~685~102 & MYTHOS & \tick & \cite{jtOH}\\
LXII & \ce{C3} & 3 & 3000 & 1 & 6~797~725~272 & AtLast & \tick & \cite{jtC3}\\
LXIII & HDO & 1 & 3000 & 1 & 3~000~000 &  \HDOname& \tick & \cite{jtHDO}\\
LXIV & PN & 1 & 5000 & 1 & 634~243 & PaiN & \tick & \cite{jt954}\\
\hline
\end{tabular}
{
\flushleft\noindent
Paper: Number in series published in Mon. Not. R. Astron. Soc.; $N_{\rm iso}$:  the number of isotopologues considered; $T_{\rm max}$: maximum temperature for which the line list is complete;  $N_{\rm elec}$: number of electronic states considered; $N_{\rm lines}$: number of lines, the value is for the main  (parent) isotopologue. \\
$^a$ The Larner line list \cite{jt447} for H$^{13}$CN/HN$^{13}$C is recommended. 
}
}
\end{table}

\subsection{MARVELization}
\label{ss:MARVEL}
In terms of the resolving power, defined via wavelength $\lambda$ and its uncertainty $\Delta \lambda$ as \(R=\frac{\lambda}{\Delta \lambda}\), calculated ExoMol line
lists are generally of sufficient accuracy to be useful for $R<10~000$. This is sufficient for current spaceborne observations of exoplanets, for example JWST has a
maximum $R=3000$. However, ground based observations, and in particular high resolution Doppler-shift cross-correlation
spectroscopy, work at much higher resolutions, typically  $R=100~000$ or even higher. It is clear that the standard theoretical method, even using empirical data to improve the
spectroscopy model \cite{jt511}, struggles to approach the level of accuracy required for this resolution. A different approach has therefore been adopted based on the explicit use of empirically
derived energy levels.

To provide empirical energy levels we used the MARVEL algorithm \cite{jt412}. Originally developed to improve representations
of water vibration-rotation spectra \cite{jt454,jt562}, the MARVEL methodology is actually agnostic about the type of spectra being studied and we have applied it widely to
rovibronic problems as well as, for VO, to hyperfine-resolved spectra \cite{jt869}. In essence MARVEL takes a list of assigned high resolution transitions with uncertainties
and inverts them to give a list of empirical energies with associated uncertainties. MARVEL does this by constructing a spectroscopic network \cite{11CsFuxx.marvel} composed of (all available) assigned,
measured transitions with uncertainties; this network is inverted using the so-called X-matrix method \cite{76FlCaMa.MARVEL} to  give empirical energy levels each with an associated uncertainty. 

The ExoMol data structure gives a single set of energy levels in the \texttt{.states} file and a set of  Einstein A coefficients in the transitions (\texttt{.trans}) file, which are processed together to compute transition wavenumbers and intensities.
This structure lends itself to straightforward improvement by replacing calculated energy levels with empirical ones, increasing the accuracy of predicted transition wavenumbers.
This approach has the advantage over simply replacing computed transition wavenumbers with measured ones because it produces accurate predictions for the transition wavenumbers of many yet to be observed transitions.
For example, Al-Derzi {\it et al.} \cite{jt828} performed a MARVEL project for formaldehyde (H$_2$CO) which used a network of 16~403 unique transitions to determine 5029 energy levels.
Substituting these 5029 energies into the AYTY ExoMol H$_2$CO line list \cite{jt597} resulted in 367~779 transitions whose wavenumbers were determined using these empirical energies.
We refer to this process as MARVELization.
In practice there are a number of ways of improving on the energy levels provided by the variational calculations as discussed by McKemmish {\it et al.} \cite{jt948}.
Even so, it is usually only possible to MARVELize a small proportion of the  energy levels (for example the AYTY line list contains over 12 billion transitions) although in general the MARVELized levels include the ones involved
in the strongest transitions.
We note that the A in MARVEL stands for active which means that MARVEL datasets can be actively updated when new high resolution measurements become available. Thus, for example, a new  optical frequency comb Fourier transform spectrum of formaldehyde
by Germann {\it et al.} \cite{jt906} was recently used to update the MARVEL results of Al-Derzi {\it et al.} \cite{jt828} and
the AYTY line list; use of this very high resolution spectrum added a further 82 new energy levels to the MARVEL compilation and,
perhaps more importantly, substantially improved the accuracy of many of the energy levels.
 Line lists that have been MARVELized are identified by \tick\ marks in Table~\ref{tab:exomoldata}.
 
To use the uncertainties in the energy levels it is assumed that the uncertainty in the transition wavenumber, $\Delta \omega$, is given by
\begin{equation}
    \Delta \omega= \sqrt{(\Delta E^\prime)^2 + (\Delta E^{\prime\prime})^2},
\end{equation}
where $\Delta E^\prime$ and $\Delta E^{\prime\prime}$ are the uncertainties in the upper and lower state energy levels, respectively. 
Note that on occasion this uncertainty will be an overestimate as there are situations where a precisely determined transition has been measured between two states whose energies are both less well determined.
However, we suggest that the current implementation should be adequate for most practical purposes.
As part of this data release we have ensured that all ExoMol-generated recommended datasets contain uncertainties in their energy levels and the source of these energies are marked using the codes given below in Sect.~\ref{S:format}.
For cases where these levels have not been MARVELized, these uncertainties can be quite large. Conversely, transitions between two states with MARVELized energies should be accurate enough to use in high resolution studies.
Our post-processing codes, \textsc{ExoCross} \cite{jt708} and \textsc{PyExoCross} \cite{jt914}, give the option of generating spectra or cross sections only using transitions with low uncertainties and our new web portal ExoMolHR, see Sect.~\ref{ss:newweb}, provides a database of high resolution transitions which can be interrogated interactively.

An underlying assumption of the MARVELization process is that the relatively small shifts in the line positions
do not significantly influence the line intensities. In most cases this is probably a reasonable assumption
but there is one set of circumstances where things are more complicated. Resonances caused by the accidental
interaction of levels in different vibrational or vibronic states can lead to severe perturbations of the
intensities caused by so-called intensity stealing between transitions. This effect is known
to be very sensitive to the details the calculation; Lodi and Tennyson \cite{jt522} designed a method
for identifying those transitions which are sensitive to these interactions by performing repeat calculations with different potential energy and dipole moment surfaces. This method has proved important for high accuracy
calculations \cite{jt625}. However, the Lodi-Tennyson method requires the generation of, at least, four line lists
for given species and thus far has not been employed for the large and usually expensive ExoMol line lists.

\subsection{Diatomics}

\subsubsection{AlO, paper XVIII}
The ATP line list for aluminium monoxide, \ce{AlO}, has been updated using MARVEL \cite{jt835}. AlO has been detected in a number of objects such as hot Jupiter exoplanets WASP-43b \cite{20ChMiKa.AlO} and HAT-P-41b \cite{21ShWeMa.exo}, as well as in eruptive young stellar objects (YSOs) \cite{jt928} using the ATP line list. The ATP line list has also been used in plasma studies \cite{17BoMaRu.AlO,17HoMaZo.AlO,23RiHuYo.AlO}.

\subsubsection{\texorpdfstring{\ce{NO}}{NO}, papers XXI and XLII}
The full rovibronic  $^{14}$N$^{16}$O XABC line list \cite{jt831} replaces the ground state NOName \cite{jt686} line list.
A MARVEL study was undertaken as part of the XABC line list construction.
However, NOName should still be used for the  minor isotopologues of NO.

\begin{figure}
\centering
\includegraphics[width=0.7\textwidth]{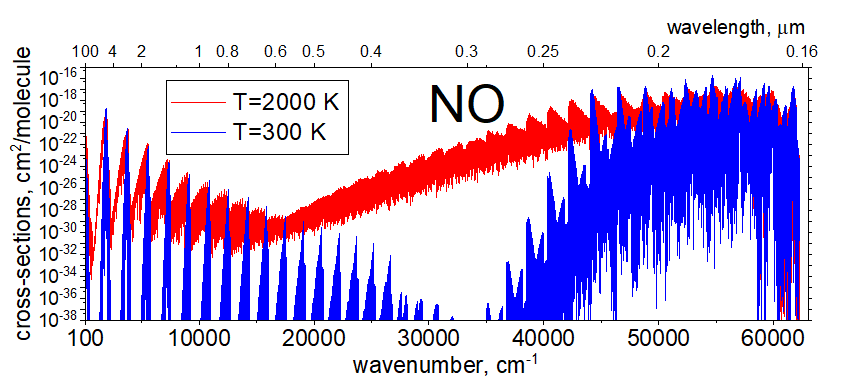}
\caption{\label{f:NO} Cross sections generated using the ExoMol line list XAB for \ce{NO} at $T=300$~K and 2000~K.}
\end{figure}

The XABC line list gives thorough coverage of the so-called $\gamma$, $\beta$ and $\delta$  band systems, which correspond, respectively, to the A\,$^2\Sigma^+$ -- X\,$^2\Pi$, B\,$^2\Pi$ -- X\,$^2\Pi$ and C\,$^2\Pi$ -– X\,$^2\Pi$ electronic bands. We note that the $\gamma$ band system has recently been proposed as a potential biomarker in exoplanetary atmospheres
\cite{23TsBi.NO}. Cross sections of \ce{NO} computed using XAB are shown in  Fig.~\ref{f:NO}.

\subsubsection{\texorpdfstring{\ce{AlH}}{AlH}, papers XXI and XLIV}
\label{subsubsec:AloHa}
Pavlenko et al. \cite{jt874} used the AlHambra line list \cite{jt732} to analyse the spectrum of \ce{AlH} in Proxima Cen. While the 
line list reproduced the majority of the observed spectrum well, it did not reproduce the predissociative lines accurately either in
their position or their widths. As a result it was decided to implement procedures \cite{jtpred} and data structures \cite{jt898} to represent line broadening due to predissociation  within the project. A new line list,  AloHa \cite{jt922}, superseded the  AlHambra line list. AloHa has been MARVELized for both AlH and AlD, and also provides an accurate representation of the predissociation lifetime. The line list also provides continuum absorption cross sections describing photoabsorption, i.e. absorption to  continuum (unbound) states. The structure of the continuum absorption cross sections is introduced in Sect.~\ref{ss:xsec}.

\subsubsection{MgO, paper XXXII}

The LiTY line list \cite{jt759} for magnesium monoxide, \ce{MgO}, has been updated for the four lowest electronic states using MARVEL \cite{jt948}. Additionally, the Predicted Shift (PS) methodology was used to improve the calculated energies and estimate energy uncertainties \cite{jt948}. The updated line list was used to update the line lists for the $^{24}$Mg$^{17}$O, $^{24}$Mg$^{18}$O, $^{25}$Mg$^{16}$O, and $^{26}$Mg$^{16}$O isotopologues with the isotopologue extrapolation procedure (IE) \cite{jt948}. MgO has been suggested as one of the constituents of the so-called lava planets \cite{22ZiVaMi.exo, 24FaTrCh.MgO}. 

\subsubsection{TiO, paper XXXIII}

The Toto line list for TiO \cite{jt760} has been widely used. Notably ToTo was successfully used by Pavlenko et al. \cite{jt799} to extract
Ti isotopic abundance using band heads of $^{50}$TiO, $^{49}$TiO, $^{48}$TiO, and $^{47}$TiO in the spectra of M dwarfs.
This work provides
endorsement of the isotopologue extrapolation procedure (IE) \cite{jt948} used to improve the line lists for minor isotopologues by using data from the main isotopologue.  The ToTo line list has recently been updated using new MARVEL energy levels derived incorporating recent 
laboratory measurements \cite{jt948}. 
The \texttt{.states} file was updated using the Predicted Shift (PS) methodology \cite{jt948}.
ToTo has been used for the detection of TiO in a number of exoplanets, such as the ultra-hot Jupiter WASP-189b \cite{22BiHoKi.exo}, WASP-33b \cite{21SeNuMo.exo}, WASP-76b \cite{20EdChBa.exo} and WASP-77Ab \cite{24EdChxx.exo}.

\subsubsection{SH, paper XXXVI}

The GYT linelist \cite{19GoYuTe} remains as the recommended ExoMol linelist for the mercapto radical and is recommended for all applications \cite{jt810}. As in hydroxyl (sec. \ref{subsection:hydroxyl}), SH exhibits strong and complex predissociation due to spin orbit interactions between the A $^2\Sigma^+$ state and the dissociative, 1~$^2 \Sigma^-$, 1~$^4 \Sigma^-$, and 1~$^4 \Pi$ states. SH/SD predissociation lifetimes will be released soon \cite{jtpred} and lifetimes in the GYT linelist will be updated as per Sect.~\ref{s:prediss}.  

\subsubsection{NaO, paper XLIII}

A new line list for \ce{NaO} called NaOUCMe \cite{jt854} has been provided; there is very limited high resolution spectroscopic data on NaO
so the line list is not suitable for high resolution studies.

\subsubsection{SiO, paper XLIV}

\begin{figure}
\centering
\includegraphics[width=0.7\textwidth]{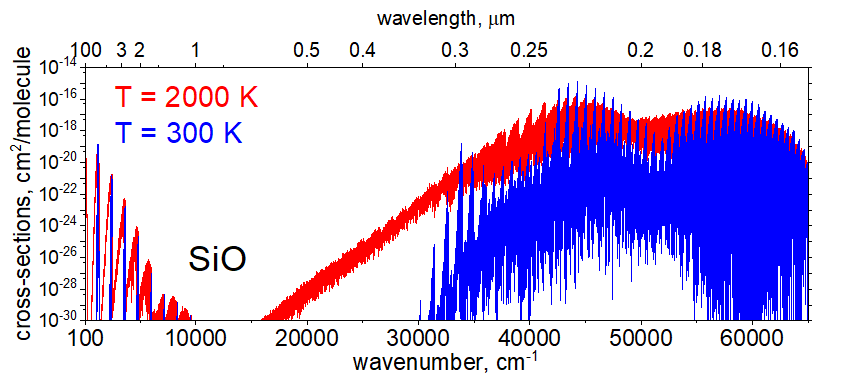}
\caption{\label{f:SiO} Cross sections generated using the ExoMol line list SiOUVenIR for \ce{SiO} at $T=300$~K and 2000~K.}
\end{figure}

A new line list for silicon monoxide, SiO, called  SiOUVenIR, has been provided \cite{jt847} and replaces the well-used EJBT rovibrational line list \cite{jt563}. The new line list has been MARVELized and also provides much broader wavelength coverage as it also covers rovibronic transitions. Like MgO, SiO has been suggested as one of the important sources of opacity in lava planets as well as sub-Neptunes \cite{15ItIkKa.exo,22NgCoPi.exo,22ZiVaMi.exo,23BoOwSc.exo,23ZiMiBu.exo,23PiGaBr.exo,24FaTrCh.MgO}. Cross sections of \ce{SiO} computed using SiOUVenIR are shown in  Fig.~\ref{f:SiO}.

\subsubsection{CaH and MgH, paper XLV}

\begin{figure}
\centering
\includegraphics[width=0.7\textwidth]{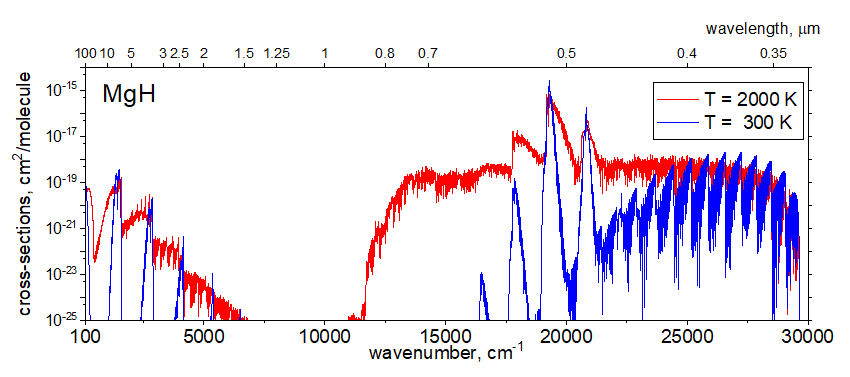}
\caption{\label{f:MgH} Cross sections of \ce{MgH} generated using the ExoMol line list XAB at $T=300$~K and 2000~K.}
\end{figure}

New rovibronic line lists for CaH and MgH called XAB have been computed \cite{jt858}. The line lists have been MARVELized and consider
the important isotopologues of both species. The line lists replace the more restricted rovibrational Yadin line lists  \cite{jt529}.
As MARVELized line lists for BeH, as well as BeD and BeT, are also available \cite{jt722}, the Yadin linelists \cite{jt529} for these alkaline earth monohydrides are no longer recommended. Cross sections of \ce{MgH} computed using the XAB line list are shown in  Fig.~\ref{f:MgH}.

\subsubsection{VO, paper XLV}

The new HyVO line list for Vanadium monoxide \cite{jt923} replaces the previous ExoMol VOMYT line list \cite{jt644}.
VOMYT was built on a relatively crude theoretical model and known to be inadequate for high resolution studies \cite{22deKeSn.VO}.
The spectrum of VO shows significant splittings due to the hyperfine structure induced by the high spin ($I=\frac{7}{2}$) and large magnetic moment of the $^{51}$V nucleus.
To create a VO line list suitable for high resolution studies, it was necessary to perform a MARVEL study which explicitly considered hyperfine effects where possible \cite{jt869}, extend our diatomic variational nuclear motion code \textsc{Duo}
to include hyperfine effects \cite{jt855} and then perform extensive model building studies \cite{jt873,jt912}.
HyVO is a fully hyperfine-resolved line list which contains nearly 59 billion transitions, a huge number for a diatomic molecule.
The new HyVO line list has been used to detect VO in the atmosphere of WASP-76b via high-resolution cross-correlation spectroscopy \cite{24MaGiNu.exo}, strengthening the findings of the previous study of this planet which utilised the old VOMYT line list \cite{23PeBeAl.exo}.

\subsubsection{SO, paper LVI}

The new SOLIS line list for the main sulfur monoxide isotopologue  $^{32}$S$^{16}$O has been computed and MARVELized \cite{jt924}. The electronic structure of SO exhibits non-adiabatic behavior where an earlier study \cite{jt876} builds a quasi-diabatic spectroscopic \emph{ab initio} model before refinement to MARVEL data. 

SO is expected to be abundant within exoplanets whose atmospheres undergo warm chemistry or have regions with prolific photochemical reactions. It was recently detected in a JWST MIR transmission spectrum of WASP-39b, a Saturn-mass exoplanet, using the SOLIS line list \cite{24PoFeLe.exo}.

\subsubsection{SiN, paper XLVI}

A new line list for silicon mononitride, SiN, called Sinful, has been computed \cite{jt858} for the main isotopologues,$^{28}$Si$^{14}$N, $^{29}$Si$^{14}$N, $^{30}$Si$^{14}$N, $^{28}$Si$^{15}$N. The line list includes 6 lowest-lying electronic states.  The line list has been MARVELized and is in good agreement with previously reported spectra. SiN is considered to be a good indicator for planet-metallicity \cite{11BrDoCo.SiN} and has been detected in different media in space \cite{22BoRiBu.SiN,03ScLeMe.SiN, 77TuDaxx.SiN, 83FeMaBe.SiN,47DAxxxx.SiN,52GRxxxx.SiN,92TUxxxx.SiN}.

\subsubsection{OH, paper LXI}\label{subsection:hydroxyl}

The new MYTHOS OH line list \cite{jtOH} replaces the one provided by MoLLIST \cite{18YoBeHo.OH} as the recommended ExoMol line list. 
The rovibronic line list considers transitions within the ground X~$^2\Pi$ electronic state to the A~$^2\Sigma^+$, B~$^2\Sigma^+$ and C~$^2\Sigma^+$ states which all support bound states. 
The model also includes four dissociative states, 1~$^2 \Sigma^-$, 1~$^4 \Sigma^-$, 1~$^4 \Pi$, and 1~$^2 \Delta$, the first three of which which causes significant predissociation broadening of levels within the  A~$^2\Sigma^+$ state \cite{jt933}; the lifetime broadening effects are explicitly allowed for in the line list. 
Transitions from the ground state to the dissociative 1~$^2 \Sigma^-$ and 1~$^2 \Delta$ states are also calculated here and are included in the temperature dependent photoabsorption cross sections. 
The line list is of a high accuracy having used the empirical energies from a previous MARVEL study \cite{jt868}. Two datasets are made available: a state bound to bound line list and temperature dependent photoabsorption cross sections which include the continuum transitions. Temperature-dependent photodissociation cross sections for OH are currently being computed.

\subsubsection{\texorpdfstring{\ce{PN}}{PN}, paper LXIV}

A rovibrational line list for phosphorus nitride, named YYLT, was computed early in the ExoMol project \cite{jt590}.
A new, MARVELized robvibronic line list PaiN \cite{jt954} covering the A--X band system and providing  improvements for the X--X transitions has been constructed  starting from the \emph{ab initio} spectroscopic model of Semenov {\it et al.} \cite{jt842}.
PaiN has replaced YYLT as the recommended line list. We note the PN is likely to be hard to detect in exoplanet atmospheres
because it has a strong vibrational fundamental in the 10 $\mu$m region, which is too long wavelength for current observational
capabilities, and the  A--X band system lies in the UV around 250 nm which is also hard to observe and unlikely to be important around cool stars.
 
\subsubsection{Other new  diatomic line lists}

New line lists are provided for \ce{AlCl} (YNAT \cite{jt887}),  \ce{CH+} (PYT \cite{jt913}) and 
\ce{YO} (AloHa \cite{jt921}),
all of which include rovibronic transitions as well as rotation-vibration transitions.

\subsection{Triatomics}

\subsubsection{\texorpdfstring{\ce{H3+}}{H3+}, Papers XX and L}

Bowesman et al. \cite{jt890} updated the MiZATeP \ce{H3+}  line list  \cite{jt666} and the ST line list for H$_2$D$^+$ \cite{jt478} using MARVEL.
New MiZo line lists for \ce{D2H+} and \ce{D3+} were generated \cite{jt890}; the MiZo \ce{D2H+} line list also uses MARVEL
energy levels while the energy levels of the  more
limited \ce{D3+} line list were improved using effective Hamiltonian data. All the line lists are suitable for high resolution studies at 
long wavelengths. A review of the role of \ce{H3+} in astrophysics has been provided by
Miller {\it et al.} \cite{jt800}.

\subsubsection{Water, Papers XIX, XXX and LXIII}

POKAZATEL \cite{jt734} remains the recommended line list for H$_2$$^{16}$O, it has been widely and successfully used.
Water is of course widely observed in exoplanet spectra and several other line lists including BT2 \cite{jt378}, HITEMP 2010 \cite{jt480} which is based on BT2, and the pioneering but now dated Partridge and Schwenke \cite{97PaScxx.H2O}. We strongly recommend the adoption
of POKAZATEL over these older, less complete and less accurate line lists for this key molecule. Indeed a near infrared
laboratory high temperature ($T=1723$~K) comparison suggested that POKAZATEL reproduced the observed spectrum
roughly six times better than BT2 \cite{20MeSaNa.H2O}.

Isotopologues H$_2$$^{17}$O and H$_2$$^{18}$O are represented by the HotWat78 line lists \cite{jt665}. While POKAZATEL 
is very complete and should give a good representation of water absorption up to at least 5000~K, the Hotwat78 line lists
are only complete up to 3000~K. We plan to extend the temperature range covered for H$_2$$^{17}$O and H$_2$$^{18}$O.

All H$_2$O  line lists discussed above have been MARVELized \cite{jt795,jt817} and are therefore suitable for high accuracy studies.

A new line list \HDOname\ has been computed for deuterated water, HDO \cite{jtHDO}. This replaces the old VTT line list \cite{jt469}. \HDOname\
is more complete than VTT. In particular, the potential energy surface (PES) used in the line list
calculations was obtained by fitting to the data \cite{jt830} up to
35~000 \cm.  Energies up to 41~000 \cm\ were used in 
the linelist calculations. 
 Unlike VTT, the  \HDOname\ energy levels have been MARVELized using energies from an IUPAC study \cite{jt482}; we note
that a MARVEL project to update these energy levels is in progress which will lead to the release of an updated version of \HDOname. 
 The spectrum of HDO is significantly different from that of H$_2$O and the 
 \HDOname\ line list should be suitable for use in high resolution studies.

In addition, room temperature line lists for the isotopologues H$_2$$^{14}$O \cite{jt942}, H$_2$$^{15}$O \cite{jt920} and H$_2$$^{19}$O \cite{jt945} have recently been added to the databases. These isotopologues, in which the radioactive oxygen atom has a half-life
in the region of a minute, may form in thunderstorms \cite{jt843}, in water cooled nuclear and fusion reactors, and as a result of using isotopes in medical procedures.

\subsubsection{\texorpdfstring{\ce{CO2}}{CO2}, Paper XXXIX}

\begin{figure}
\centering
\includegraphics[width=0.7\textwidth]{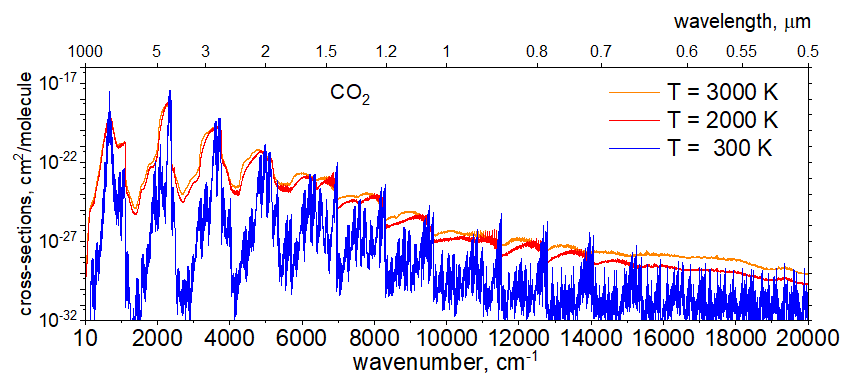}
\caption{\label{f:CO2} Cross sections generated using the ExoMol line list UCL-4000 for \ce{CO2} at $T=300$~K, 2000~K and 3000~K.}
\end{figure}

The UCL-4000 line list \cite{jt804} for carbon dioxide has been produced an accurate PES AMES-1 \cite{12HuScTa.CO2} and a high-level \ai\ dipole moment surface \cite{jt613} of \ce{CO2} using the TROVE variational program \cite{TROVE_prog}.  UCL-4000 has been widely used but suffers from two issues. It is only for the parent, $^{12}$C$^{16}$O$_2$, isotopologue and at visible wavelengths it was found to overestimate transition intensities at higher wavenumbers, a problem found with other hot CO$_2$ line lists \cite{24BaRaWo.CO2}. We are therefore in the process of constructing an improved hot line list for this important molecule which will consider all important isotopologues and be MARVELized \cite{jt925,jt932}. Cross sections of \ce{CO2} generated using the UCL-4000 line list are illustrated in Fig.~\ref{f:CO2}. 

The same methodology used for UCL-4000  was subsequently used to produce an electric quadrupole (E2) line list for \ce{CO2} \cite{21YaCaYu}. The theoretical E2 transitions were validated experimentally using in the 3.3~$\mu$m region \cite{21FlGrMo.CO2} and used to detect  E2 lines in the Martian atmospheric spectra from the ExoMars ACS observations \cite{18KoBeDo.planets}.

\subsubsection{LiOH, NaOH, KOH, Papers XLI and LI}

\begin{figure}
\centering
\includegraphics[width=0.7\textwidth]{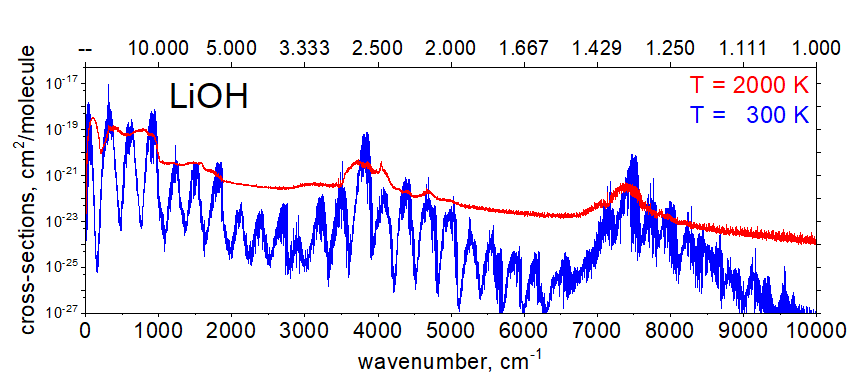}
\caption{\label{f:LiOH} Cross sections of LiOH generated using the ExoMol line list OYT7 at $T=300$~K and 2000~K.}
\end{figure}

New line lists for the alkali metal hydoxides KOH, NaOH and LiOH named  OYT4 \cite{jt820}, OYT5 \cite{jt820} and OYT7 \cite{jt905}, respectively,
have been calculated. There is a lack of high resolution spectroscopic data for these systems so these are not MARVELized and therefore not
suitable for high resolution studies.
Cross sections of \ce{LiOH} generated using the OYT7 line list are illustrated in Fig.~\ref{f:LiOH}. 

\subsubsection{CaOH, Paper XLVII}

\begin{figure}
\centering
\includegraphics[width=0.7\textwidth]{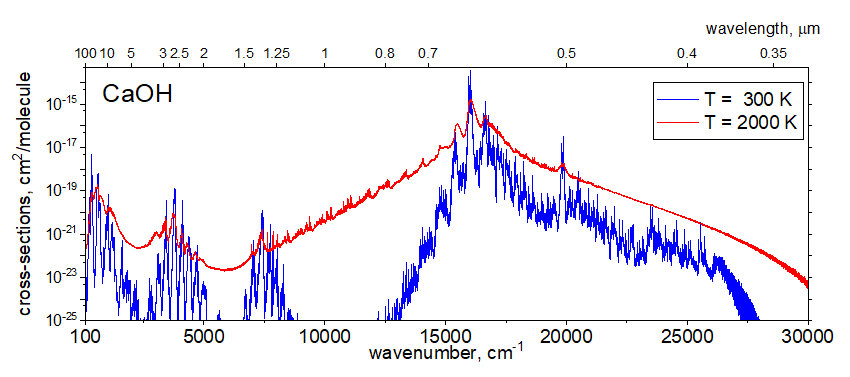}
\caption{\label{f:CaOH} Cross sections of \ce{CaOH} generated using the ExoMol line list OYT6 at $T=300$~K and 2000~K.}
\end{figure}

A new line list called OYT6 is provided for calcium monohydroxide, CaOH \cite{jt858}. This is the first rovibronic line list for a triatomic molecule which
includes consideration of the important $\tilde{A}\,^2\Pi$ -- $\tilde{X}\,^2\Sigma^+$ electronic band system as well as ground state
rotation-vibration transitions.  This line list was built on a MARVEL study \cite{jt791}, then constructed a spectroscopic model \cite{jt838} using the  rovibronic nuclear motion code for triatomics  \textsc{EVEREST} \cite{EVEREST}. CaOH is known to be an important
component of the atmosphere of cool stars \cite{00ReKiGi.CaOH}; the OYT6 line list has been used as a thermometer in combustion and explosion applications \cite{23Fateev.CaOH}.
Cross sections of \ce{CaOH} computed using the OYT6 line list are shown in  Fig.~\ref{f:CaOH}.

\subsubsection{OCS, Paper LVIII }

A new line list called OYT8 is provided for carbonyl sulphide, OCS \cite{jt943}; this line list uses a previous MARVEL study \cite{jt916} and therefore
should be suitable for high resolution studies. Given that CO \cite{23EsLoAd.CO}, CO$_2$ \cite{23AhAlBa.exo}, SO$_2$ \cite{23TsLePo.exo,24PoFeLe.exo}, and now SO 
have been detected by JWST in exoplanetary atmospheres, the presence of OCS would also appear likely. Major studies are already searching for its spectroscopic signature \cite{23TsLePo,23AlWaAl.exo}.

\subsubsection{\texorpdfstring{\ce{N2O}}{N2O}, Paper LIX}
A new line list for nitrous oxide, \ce{N2O}, has been computed \cite{jt951}.  \ce{N2O} is prominent in the Earth's atmosphere and considered to be an observable potential biomarker on exoplanets \cite{17Grenfe,22ScOlPi.N2O,24AnPiLe.N2O}. Five isotopologues are considered, $^{14}$N$_2$$^{16}$O, $^{15}$N$^{14}$N$^{16}$O, $^{14}$N$^{15}$N$^{16}$O, $^{14}$N$_2$$^{17}$O and $^{14}$N$_2$$^{18}$O. 
The parent isotopologue $^{14}$N$_2\,^{16}$O was MARVELized using the MARVEL energies of Tennyson {\it et al.} \cite{jt908}.
MARVEL studies on the four  isotopologues arising from substituting a single atom are in progress.

\subsubsection{\texorpdfstring{\ce{C3}}{C3}, Paper LXII}

A new line list for \ce{C3} has been calculated, which considers ro-vibrational transitions within the ground state. Three isotopologues are considered: $^{12}$C$_3$, $^{12}$C$^{13}$C$^{12}$C and the asymmetric $^{12}$C$^{12}$C$^{13}$C. The $^{12}$C$_3$ line list was MARVELized using results from
a previous MARVEL study \cite{jt915}.

\subsection{Tetratomics}

\subsubsection{\texorpdfstring{\ce{H2CO}}{H2CO}, Paper VIII}

\begin{figure}
\centering
\includegraphics[width=0.7\textwidth]{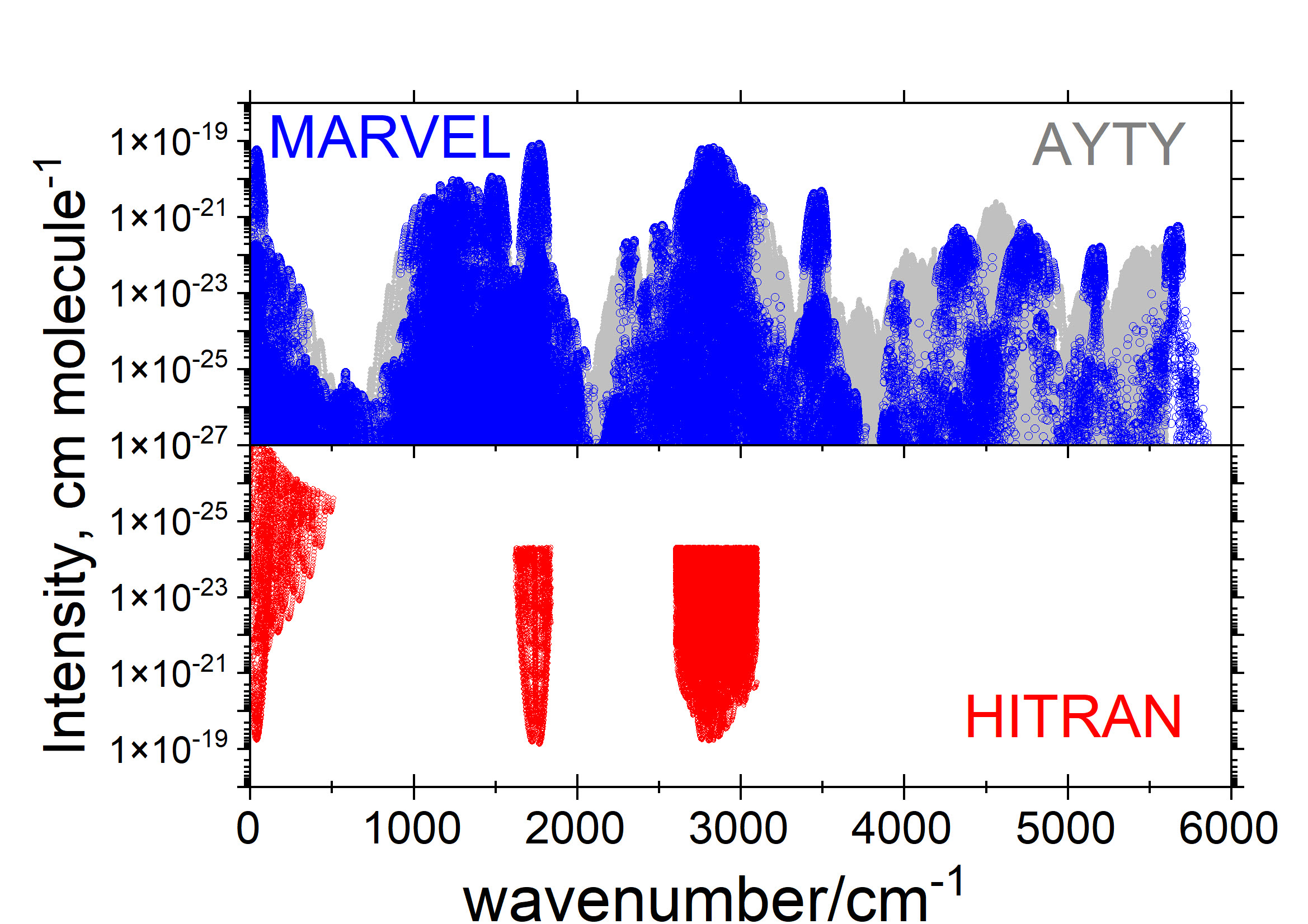}
\caption{\label{f:H2CO} Room temperature ($T = 296$ K) spectra of \ce{H2CO} from three different sources: lower panel, HITRAN; upper panel, MARVELized transitions \cite{jt828,jt906} overlaid with  the AYTY line list \cite{jt597}.}
\end{figure}

The AYTY line list for formaldehyde \cite{jt597}, H$_2$$^{12}$C$^{16}$O, has been updated using MARVEL  \cite{jt828,jt906} by replacing the calculated AYTY energy levels with MARVEL ones. Based on this update, 373~160  transition frequencies with experimental accuracy were determined. These are illustrated in Fig.~\ref{f:H2CO}, where they are compared to 40~670 transitions of H$_2$$^{12}$C$^{16}$O in HITRAN~2020. A line list for H$_2$$^{13}$CO is under construction.

\subsubsection{\texorpdfstring{\ce{NH3}}{NH3}, Papers XXXV and LX}

The CoYuTe line list for  $^{14}$NH$_3$ ammonia \cite{jt771} has been updated using new  MARVEL data \cite{jt784}.
The CoYuTe line list has been and is continuing to be used to assign ammonia spectra in the near-infrared and to analyze plasma emissions spectra \cite{24SaFlHa.NH3}. Laboratory spectra
of ammonia need to be analyzed in this region and into the visible not least because pronounced ammonia features
are observed in Jupiter in the near-infrared and visible. In particular, the feature at 6474~\AA\  is modelled by  CoYuTe \cite{jt743,jt745} better than other
available compilations but for which CoYuTe still shows a noticeable shift  in wavelength compared to the observations.

A new $^{15}$NH$_3$ line list called   CoYuTe-15 was generated using the same procedure as for the $^{14}$NH$_3$ CoYuTe line list \cite{jt771}, including
MARVELization; it replaces the previous room-temperature  $^{15}$NH$_3$ line list  \cite{15Yurche.NH3} which is also less accurate and
not MARVELized. The recent detection of $^{15}$NH$_3$  in the atmosphere of a cool brown dwarf \cite{23BaMoPa.dwarfs} suggests that this
isotopologue could also be detected in the atmospheres of exoplanets.

\subsubsection{HCCH, Paper XXXVII}

The  aCeTY line list for acetylene, \ce{C2H2}, has been updated using MARVEL \cite{jt705}.

\subsubsection{\texorpdfstring{\ce{H2CS}}{H2CS}, Paper XLVIII}

\begin{figure}
\centering
\includegraphics[width=0.7\textwidth]{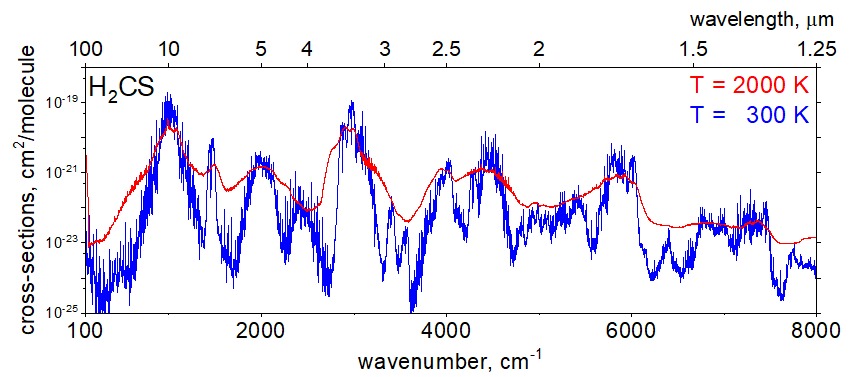}
\caption{\label{f:H2CS} Cross sections generated using the ExoMol line list MOTY for \ce{H2CS} at $T=300$~K and 2000~K.}
\end{figure}

A new line list for thioformaldehyde, \ce{H2CS}, called MOTY has been computed by Mellor {\it et al.} \cite{jt886}; the energy levels were on the basis of a  MARVEL study \cite{jt885} to make it suitable for high resolution studies. Figure~\ref{f:H2CS} illustrates the cross sections of \ce{H2CS} computed using MOTY. 

\subsection{Pentatomics}

\subsubsection{\texorpdfstring{CH$_4$}{CH4}, Paper LVII}

\begin{figure}
\centering
\includegraphics[width=0.7\textwidth]{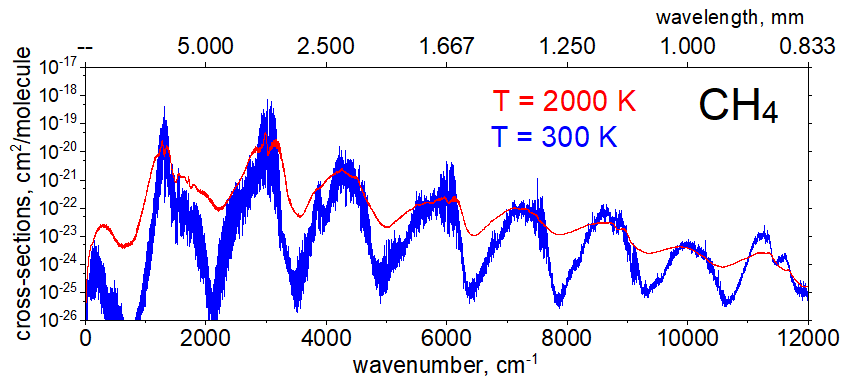}
\caption{\label{f:CH4} Cross sections of \ce{CH4} generated using the ExoMol line list MM at $T=300$~K and 2000~K.}
\end{figure}

A completely new line list for methane, \ce{CH4}, called MM \cite{jt927} 
replaces the previous ExoMol methane line lists 10to10 \cite{jt564} and 30to10 \cite{jt698}. MM is based on the use of
improved theory, an improved potential energy surface plus extensive MARVELization based on a comprehensive, parallel MARVEL study \cite{jt926}.
Line lists for hot methane are also available from Theorets \cite{14ReNiTy.CH4,17ReNiTya.CH4} and HITEMP \cite{20HaGoRe.CH4}.
MM is broadly comparable with the Theorets line list but
at high temperatures ($>1000$~K)  is significantly  more complete than the HITEMP line list. Cross sections of \ce{CH4} generated using the MM line list are illustrated in Fig.~\ref{f:CH4}.

There are many unassigned or partially assigned published high resolution spectra of methane available \cite{jt926}. We are actively
using MM to assign these spectra \cite{jtCH4}. Any new assignments we make, or indeed new high resolution spectra, will be used in an extended
MARVEL study, the results of which can be used to re-MARVELize and further improve the accuracy of the MM line list.

\subsection{Other line lists}
\label{ss:other}

\begin{table}[H]
\centering
\caption{Datasets recommended as part of the ExoMol project but imported from other sources. } 
\label{tab:otherdata}
\small
\begin{tabular}{lcrcrlcll}
\hline
Molecule&$N_{\rm iso}$&$T_{\rm max}$&$N_{\rm elec}$&$N_{\rm lines}$&DSName&&Reference&Methodology\\
\hline
CH      &       1       &5000&  4       &       52201   &       MoLLIST& &\cite{14MaPlVa.CH}&Empirical \\
AlF     &               1       &5000   & 1     &       40~490   &       MoLLIST& &\cite{18YoBexx.AlF}&Empirical \\
CaF     &       1&5000  & 6             &       14~817   &       MoLLIST& &\cite{18HoBexx.CaF}&Empirical \\
MgF     &1      &       5000& 3 &       8136    &       MoLLIST& &\cite{17HoBexx.MgF}&Empirical \\
KF      &1 &    5000&   2       &       10~572   &               MoLLIST& &\cite{16FrBeBr.NaF}&Empirical \\
NaF     &       1 &5000 &       1       &       7884    &               MoLLIST& &\cite{16FrBeBr.NaF}&Empirical \\
LiCl    &       1&      5000& 4 &       26~260   &               MoLLIST& &\cite{18BiBexx.LiF}&Empirical \\
LiF     &       1       & 5000& 2       &       10~621   &       MoLLIST& &\cite{18BiBexx.LiF}&Empirical \\
TiH     &       1 &     5000&   3       &       181~080  &       MoLLIST& &\cite{05BuDuBa.TiH}&Empirical \\
CrH     &       1&      5000&2  &       13~824   &               MoLLIST& &\cite{06ChMeRi.CrH} &Empirical\\
FeH     &       1 &5000 &       2               &       93~040   &       MoLLIST& &\cite{10WEReSe.FeH} &Empirical\\
LaO     &       1 &8000 &       2               &       2~066~535   &     BDL& &\cite{22BeDoLi.LaO} &Empirical\\
HF      &       2& 5000& 1 &    7956    &  Coxon-Hajig  & &\cite{13LiGoLe.HCl,15CoHaxx.HF} &Empirical\\
HCl     &       4& 5000& 1 &    2588    & HITRAN& &\cite{11LiGoBe.HCl,13LiGoHa.HCl} &Empirical\\
HBr     &       1& 5000& 1 &    3039    & HITRAN-HBr& &\cite{13LiGoHa.HCl,15CoHaxx.HF} &Empirical\\
CO      &      9  & 9000 & 1 & 125~496& Li2015& & \cite{15LiGoRo.CO} &Empirical\\
CP      &       1& 5000&        2&      28752 &               MoLLIST& &\cite{14RaBrWe.CP} &Empirical\\
ScH&1&5000&6&1~152~827&LYT& &\cite{jt599}&Ab initio\\
LiH&1&12000&1&18~982&CLT& &\cite{jt506}&Ab initio\\
LiH$^+$&1&12000&1&332&CLT& &\cite{jt506}&Ab initio\\
HeH$^+$&4&9000&1&1430&ADSJAAM& &\cite{19AmDiJo}&Ab initio\\
H$_2$&1&10000&1&4712& RACPPK&&\cite{19RoAbCz.H2}&Ab initio\\
H$_2$&1&10000&3&32~915& ARLR&&\cite{94AbRoLa}&Ab initio\\
HD&1&9000&1&10285&ADSJAAM& &\cite{19AmDiJo}&Ab initio\\
HD$^+$&1&9000&1&10285&ADSJAAM& &\cite{19AmDiJo}&Ab initio\\
CH$_3$F&1&300&1&139~188~215&OYKYT& &\cite{19OwYaKu.CH3F}&Ab initio\\
AsH$_3$&1&300&1&3~600~000&CYT18& &\cite{jt751}&Ab initio\\
P$_2$H$_2$$^a$&2&300&1&10~667~208~951&OY-Trans& &\cite{19OwYuxx.P2H2}&Ab initio\\
P$_2$H$_2$$^a$&2&300&1&11~020~092~365&OY-Cis& &\cite{19OwYuxx.P2H2}&Ab initio\\
PF$_3$&1&300&1&68~000~000~000&MCYTY& &\cite{jt752}&Ab initio\\
CH$_3$&1&1500&1&2~058~655~166&AYYJ& &\cite{19AdYaYu.CH3}&Ab initio\\
BeH&3&5000 &2&592~308&Darby-Lewis& &\cite{jt722}&ExoMol\\
SiH$_2$&1&2000&1&254~061~207&CATS & &\cite{jt779} &ExoMol\\
CN      &       4& 10000&        3&      2~285~103   &               Trihybrid&\tick&\cite{21SyMcXx.CN} &ExoMol\\
ZrO& 6& 10000& 10& 47~662~773& ZorrO & \tick&\cite{23PeTaMc.ZrO}& ExoMol\\
NH& 4& 10000& 5& 327~014& kNigHt & \tick&\cite{24PeMcxx.NH}& ExoMol\\
HBO& 2& 3000& 1 & 142~038~890& LQL &&\cite{24LiQiLi.HBO} &ExoMol\\
N$_2$   &       1&10000 & 4&        7~182~000       &       WCCRMT  &&\cite{18WeCaCr.N2} &Empirical\\
\ce{H2O}$^b$ & 1 & 300 & 1 &  109~263 & CKYKKY &&  \cite{20CaSoSo.H2O} & Empirical \\
\hline
\end{tabular}

$N_{\rm iso}$ Number of isotopologues considered;\\
$T_{\rm max}$ Maximum temperature for which the line list is complete;\\
$N_{\rm elec}$ Number of electronic states considered;\\
$N_{\rm lines}$  Number of lines: value is for the main isotope.\\
\tick\ indicates line list that provide energy levels
with individual uncertainties and contain lines suitable for use in high resolution studies.\\
$^a$ There are separate line lists for cis and trans P$_2$H$_2$.\\
$^b$ An electric quadrupole (E2) line list for H$_2^{16}$O.
\end{table}
Table~\ref{tab:otherdata} gives an overview of recommended line lists available through the ExoMol website but not generated by the ExoMol project. As for Table~\ref{tab:exomoldata}, an indication
is given as to whether the data as presented are appropriate for high resolution studies. For Table~\ref{tab:otherdata} this information may be less useful because while most of the sources do indeed  give accurate predictions for the transition wavenumbers, the line lists  do not contain uncertainties for their individual energy levels which means that within the
ExoMol data structure wavenumber uncertainties cannot be estimated. These datasets have not been marked as suitable for high resolution studies and it is necessary to check the original
(cited) reference to establish the accuracy of the transitions.

A number of the line lists presented in Table~\ref{tab:otherdata} are taken from Bernath's MoLLIST project \cite{MOLLIST} including
the one for the CP radical \cite{14RaBrWe.CP}. We note that an alternative line list for CP has been computed by Qin {\it et al.}
\cite{21QiBaLi.CP}. A new MoLLIST line list  covering the B~$^2\Sigma^+$ - X~$^2\Sigma^+$ band system of LaO by Bernath {\it et al.} \cite{22BeDoLi.LaO} has been added. 
Compared to the 2020 release a number of MoLLIST line lists, namely OH, AlCl, CaH and NH, have been replaced by ones generated
by ExoMol.

Diatomic line lists for four molecules have also been taken from HITRAN \cite{jt857}, namely those for CO, HCl, HF and HBr.
These line lists provide extensive compilations of rotation-vibration transitions within the electronic ground states and are
appropriate for elevated temperatures. In the case of CO, a number of improvements to the treatment of the already very
accurate ground state line list \cite{jt871,jt900} and extensions to consider rovibronic transitions \cite{jt938} have been
undertaken. These will be used to provide updated and greatly extended line lists.

HITRAN data have also been used to create opacities for the oxygen molecule; these data do not extend to high temperature.
An extensive, ExoMol high-temperature line list for O$_2$ should be available soon. To this end, new methodology to treat  magnetic dipole and  electric quadrupole transitions in diatomic molecules  has been developed and implemented in \textsc{Duo} \cite{21SoYuYa,24SoYuKi}.

Molecular hydrogen, H$_2$, is one of the relatively few  molecules
for which the database provides more than one recommended line list. The weak, dipole-forbidden infrared spectrum of  H$_2$ is given by the RACPPK 
line list \cite{19RoAbCz.H2} which includes electric quadrupole transitions, for which $\Delta J = 0, \pm 2$, 
and magnetic dipole transitions with $\Delta J = 0$ only. The dipole-allowed Lyman (B~$^1\Sigma^+_u$ -- X~$^1\Sigma^+_g$) and Werner (C~$^1\Pi_u$ -- X~$^1\Sigma^+_g$) H$_2$ electronic band systems are provided by the ARLR \cite{94AbRoLa}
line list. An infrared line list of HD is provided from high accuracy \emph{ab initio} calculations by Amaral {\it et al.} (ADSJAAM) \cite{19AmDiJo}; in this case it is also necessary
to consider weak dipole-allowed transitions which arise from failure of Born-Oppenheimer approximation. \emph{Ab initio} infrared 
line lists for HD$^+$ and HeH$^+$ from Amaral {\it et al.} \cite{19AmDiJo} are also included. \emph{Ab initio} line lists for LiH$^+$ and 
LiH, designed primarily for studies of the early Universe, due to Coppola {\it et al.} \cite{jt506} are also included.  

An electric quadrupole (E2) line list CKYKKY for H$_2^{16}$O has been produced \cite{20CaSoSo.H2O} using the POKAZATEL PES and a new \ai\ electric  quadrupole moment surface of \ce{H2O}. The line list was used to identify E2 transitions of H$_2^{16}$O  experimentally for the first time \cite{20CaSoSo.H2O}.

The UNSW (University of New South Wales) group of McKemmish have generated a number of ExoMol style diatomic line lists which are available and
recommended in the database. All these line lists have been the subject of an associated MARVEL study and therefore provide high accuracy data and are suitable for high resolution studies. They are:
\begin{itemize}
    \item CN: Syme and McKemmish \cite{21SyMcXx.CN} ExoMol style line list from UNSW calculated using \textsc{Duo} \cite{Duo} hybridised with MARVEL \cite{20SyMc} and MoLLIST \cite{14BrRaWe.CN} data.
    \item ZrO: Perri et al. \cite{23PeTaMc.ZrO} ExoMol style line list from UNSW calculated using \textsc{Duo} \cite{Duo} hybridised with MARVEL \cite{18McBoGo} and MoLLIST \cite{21SoBe.ZrO} data.
    \item NH: Perri and McKemmish \cite{24PeMcxx.NH} ExoMol style line list from UNSW calculated using \textsc{Duo} \cite{Duo} hybridised with MARVEL \cite{19DaShJo} and MoLLIST \cite{14BrBeWe.NH,15BrBeWe.NH,18FeBeHo.NH} data.
\end{itemize}
In addition Li {\it et al.} \cite{24LiQiLi.HBO} have recently used ExoMol methodology to compute line lists
for two isotopologues of HBO. This line list has been
added to the datebase.

Finally the website provides partial data for two other diatomic molecules, \ce{N2} and NiH.

Though the nitrogen molecule has a singlet ground state, at present data are only provided for transitions between excited triplet states; these transitions are
 prominent in nitrogen and air plasmas. The empirical
dataset WCCRMT line list due to Western {\it et al.} \cite{18WeCaCr.N2} covering the 4~500 -- 15~700 \cm\ region is available. Recently, Jans \cite{24Jans.N2} extended this work to consider  the 
N$_2$ (C~$^3\Pi_u$ -- B~$^3\Pi_g$) second positive system, providing the data in ExoMol format. It is planned to merge these two datasets to form a single   N$_2$ triplet dataset.

The website currently provides a link to a list of observed nickel monohydride (NiH) transitions recorded by Vallon {\it et al.} \cite{09VaRiCr.NiH} and Harker {\it et al.} \cite{13HaRiTo.NiH}.
These datasets which consider the isotopologues $^{58}$NiH, $^{60}$NiH and $^{62}$NiH as well as the effects of magnetic fields, are not presented in ExoMol format. However, Havalyova \cite{21HaBoPa.NiH}
has constructed a high accuracy, empirical spectroscopic model for NiH which is currently being used as the starting point for constructing an ExoMol line list for this system.

\subsection{Atoms}
\label{s:atoms}

Thus far the ExoMol database has concentrated on molecular spectra. However, atomic
spectra are also important for exoplanet studies. So far the website contains data
on two neutral atoms:\\
\begin{itemize}
    \item Potassium: Kurucz-Allard dataset for $^{39}$K \cite{16AlSpKi.K,11Kurucz.db}
    \item Sodium: Kurucz-Allard dataset for $^{23}$Na \cite{19AlSpLe.Na,11Kurucz.db}
\end{itemize}
These data are given only in the form of opacities \cite{jt801}. We note that Allard {\it et al.} \cite{24AlKiMy.atom} have recently
updated the opacities for broadening of K by He.

\subsection{Partition functions}

Partition functions are provided for all molecular line lists in two column format
of temperature in K and partition function. The data are provided in 1 K steps up
to the temperature maximum specified by the line list and given in the \texttt{.def} file. ExoMol follows the HITRAN
convention \cite{jt692} and includes the full nuclear spin degeneracy contribution in the partition
function unlike many astronomy oriented compilations of partition functions \cite{81Irwin.partfunc,84SaTaxx.partfunc,16BaCoxx.partfunc} which
use a reduced nuclear spin factor. Our convention makes the treatment of hyperfine resolved
spectra, as done in the HyVO VO line list  \cite{jt923}, straightforward.

It is possible to compute partition functions by direct summation of the levels
provided in the \texttt{.states} file; both \textsc{ExoCross} \cite{jt708} and \textsc{PyExoCross} \cite{jt914} offer
this service. However, we recommend using the partition functions provided in the .pf file
as part of the line list, as for some molecules these represent more reliable values than those will be obtained by direct summation \cite{jt571}.

\section{Data provided}
\label{ss:data}
Table~\ref{tab:datsum} provides a summary of the various datasets available for each isotopologue on the ExoMol website.
These data can be either downloaded manually or accessed through the API (applications program interface) described in Sec.~\ref{SS:API} below.

The following subsections describe some of these datasets, again concentrating mainly on changes since the 2020 release.

\begin{table}[H]
    \centering
    \caption{Summary of data provided by the ExoMol Database.} 
    \label{tab:datsum}
    \begin{tabular}{ll}
    \hline
    Data type 
    \\
    \hline
    Line lists \\
    Absorption cross sections \\
    Pressure broadening coefficients \\
    Temperature dependent super-lines (histograms) \\
    Partition functions  \\
    Cooling functions \\
    Specific heat - heat capacity \\
    Temperature and pressure dependent opacities \\
    Photo-absorption continuum cross section$^*$\\
    Photo-dissociation cross sections including VUV absorption$^*$\\
    Spectroscopic Models    \\
    \hline
\end{tabular}
{
\flushleft
\noindent
$^*$ Denotes a new or updated data type, see text for details.
}
\end{table}

\subsection{ExoMol format}
\label{S:format}

\begin{table}
    \caption{Specification of the ExoMol file types. (Contents in brackets are optional.)}
    \label{tab:files}
    \resizebox{\textwidth}{!}{
    \begin{tabular}{lcll}
    \hline
    File extension & $N_{\rm files}$&File DSname &  Contents\\
    \hline
    \texttt{.all} & 1 & Master & Single file defining contents of the ExoMol database. \\
    \texttt{.def} & $N_{\rm tot}$ & Definition & Defines contents of other files for each isotopologue.\\
    \texttt{.states} & $N_{\rm tot}$ & States & Energy levels, quantum numbers, Uncertainties, lifetimes, (Land\'e $g$-factors). \\
    \texttt{.trans} & $^a$ & Transitions & Einstein $A$ coefficients, (wavenumber). \\
    \texttt{.broad} & $N_{\rm mol}$ & Broadening & Parameters for pressure-dependent  line profiles. \\
    \texttt{.cross} & $^b$ & Cross sections & Temperature or temperature and pressure-dependent cross sections. \\
    \texttt{.kcoef} & $^c$ & $k$-coefficients & Temperature and pressure-dependent $k$-coefficients. \\
    \texttt{.pf} & $N_{\rm tot}$ & Partition function & Temperature-dependent partition function. \\
    \texttt{.cf} & $N_{\rm tot}$ & Cooling function & Temperature-dependent cooling function. \\
    \texttt{.super} & $^d$ & Super-lines & Temperature dependent super-lines (histograms) on a wavenumber grid. \\
    \texttt{.nm} & $^e$ & VUV cross sections & Temperature and pressure dependent VUV cross-sections (wavelength, nm). \\
    \texttt{.fits}, \texttt{.h5}, \texttt{.kta} & $^f$ & Opacities & Temperature and pressure dependent opacities for radiative-transfer applications. \\
    \hline
    \texttt{.overview} & $N_{\rm mol}$ & Overview & Overview of datasets available. \\
    \texttt{.readme} & $N_{\rm iso}$ & Readme & Specifies data formats.\\
    \texttt{.model} & $N_{\rm iso}$ & Model & Model specification.\\
    \hline
    \end{tabular}
    }
    \noindent
    \small
    $N_{\rm files}$ total number of possible files; \\
    $N_{\rm mol}$ Number of molecules in the database; \\
    $N_{\rm tot}$ is the sum of $N_{\rm iso}$ for the $N_{\rm mol}$ molecules in the
    database; \\
    $N_{\rm iso}$ Number of isotopologues considered for the given molecule. \\
    $^a$ There are $N_{\rm tot}$ sets of \texttt{.trans} files but for molecules with large
    numbers of transitions the \texttt{.trans} files are subdivided into wavenumber regions. \\
    $^b$ There are $N_{\rm cross}$ sets of \texttt{.cross} files for each isotopologue. \\
    $^c$ There are $N_{\rm kcoef}$ sets of \texttt{.kcoef} files for  each isotopoplogue. \\
    $^d$ There are $N_T$ sets of $T$-dependent super-lines. \\
    $^e$ There are $N_{\rm VUV}$ sets of VUV cross sections. \\
    $^f$ Set of opacity files in in the format  native  to specific radiative-transfer programs. 
\end{table}

Table~\ref{tab:files} gives an overview of the ExoMol file structure.
Note that ExoCross/pyExocross generates other file types such as \texttt{.os} for oscillator strengths but these files are not distributed via
the ExoMol website/database.

The core of the ExoMol data structure is the provision of a \texttt{.states} and \texttt{.trans} file \cite{jt548} which give, respectively, the energy levels with associated quantum numbers and Einstein A coefficients identified by upper and lower state indices (counting numbers) which point to the \texttt{.states} file. The format of the \texttt{.trans} file is unchanged from previous releases while there have been some minor changes to data structure of the \texttt{.states}. Tables~\ref{t:trans} and \ref{t:states} present the structure of these two files. The states and, particularly, the \texttt{.trans} files can be very large so both files are supplied compressed using
\texttt{.bz2} format. For larger line lists the transitions are provided as a series of files which store data for a wavenumber region specified
in the file name. Wavenumbers are only provided in the \texttt{.trans} file of smaller line list and even in these
cases it is recommended to recompute them using the energies in the  \texttt{.states} file as they may not reflect improvements to the states file through various corrections.

\begin{table}
\centering
    \caption{Transitions file specification.}
    \label{t:trans}
    \begin{tabular}{llll}
    \hline
    Field & Fortran Format & C Format & Description \\ \hline
    $i$ & \texttt{I12} & \texttt{\%12d} & Upper state ID \\
    $f$ & \texttt{I12} & \texttt{\%12d} & Lower state ID \\
    $A$ & \texttt{ES10.4} & \texttt{\%10.4e} & Einstein $A$ coefficient in
    $\mathrm{s^{-1}}$ \\
    $\tilde{\nu}_{fi}$ & \texttt{E15.6} & \texttt{\%15.6e} & Transition wavenumber in
    cm$^{-1}$ (optional). \\
    \hline
    \end{tabular}
    {
    \flushleft
    \noindent
    Fortran format: \texttt{(I12,1x,I12,1x,ES10.4,1x,ES15.6)}.\\
    }
\end{table}

\begin{table}
    \caption{Specification of the \texttt{.states} file including
    extra data options; the formats at the end of the table are for the compulsory section only.}
    \label{t:states}
    \begin{tabular}{llll}
    \hline
    Field & Fortran Format & C Format & Description \\
    \hline
    $i$ & \texttt{I12} & \texttt{\%12d} & State ID \\
    $\tilde{E}$ & \texttt{F12.6} & \texttt{\%12.6f} & Recommended state energy in $\mathrm{cm^{-1}}$ \\
    $g_\mathrm{tot}$ & \texttt{I6} & \texttt{\%6d} & State degeneracy \\
    $J$ & \texttt{I7/F7.1} & \texttt{\%7d/\%7.1f} & Total angular momentum quantum number, $J$ or $F$
    (integer/half-integer ) \\
    Unc & \texttt{F12.6} & \texttt{\%12.6f} & Uncertainty in the state energy in $\mathrm{cm^{-1}}$ \\
    $\tau$ & \texttt{ES12.4} & \texttt{\%12.4e} & State lifetime (aggregated radiative and predissociative lifetimes) in s \\
    ($g$) & \texttt{F10.6} & \texttt{\%10.6f} & Land\'e $g$-factor (optional) \\
    (QN) & \multicolumn{2}{l}{See \texttt{.def} file} & State quantum numbers, may be several columns (optional) \\
    (Abbr) & \texttt{A2} & \texttt{\%2s} & Abbreviation giving source of state energy, see Table~\ref{tab:abbr}. \\
    ($\tilde{E_0}$) & \texttt{F12.6} & \texttt{\%12.6f} & Calculated state energy in $\mathrm{cm^{-1}}$ (optional) \\
    \hline
    \end{tabular}
    {
    \flushleft
    \noindent
    Fortran format, $J$ integer:
    \texttt{(I12,1x,F12.6,1x,I6,I7,1x,F12.6,1x,ES12.4,1x,F10.6)}\\
    or $J$ half-integer: \texttt{(I12,1x,F12.6,1x,I6,F7.1,1x,F12.6,1x,ES12.4,1x,F10.6)}.\\
    }
\end{table}

A sample extract from a \texttt{.states} file is given in Table~\ref{t:MgOstates}. The \texttt{.states} file gives the state ID, which should correspond to the line number,
and is used by the \texttt{.trans} file to identify initial and final states. The state energy, degeneracy and total angular momentum are then specified.
For most species the total angular momentum corresponds to the quantum number $J$ but for hyperfine resolved line lists this is $F$ where \(F = |\underline{J}+\underline{I}|\) and $I$ is the nuclear spin. The uncertainty, introduced in the 2020 release, is now a compulsory element of 
recommended ExoMol line lists, although it is still not available for older line lists and most non-ExoMol ones. The uncertainty provides
important information for high resolutions studies and can also help to  inform  future experimental priorities. The state lifetime, $\tau$, is also now compulsory for recommended ExoMol line lists; its
definition is now generalised so that $\tau$ gives the lifetime obtained by considering both radiative decay \cite{jt624} and predissociation effects,
see Tennyson {\it et al.} \cite{jt898}.

The optional fields in the \texttt{.states} file are defined in the \texttt{.def} discussed below. These fields generally comprise a set of quantum numbers which characterize the state. These always include the total symmetry of the state, either in the form of a total parity (denoted $+/-$), a  rotationless parity (denoted e/f) or  an irreducible representation of the molecular symmetry group the molecule is associated with. The latter is more typical for polyatomic molecules with complex symmetry groups.  Other quantum numbers are always approximate; for many polyatomic molecules, two or even three sets of quantum numbers are available. For those line lists denoted as being high resolution in Table~\ref{tab:exomoldata} two extra columns have been added to  the \texttt{.states} file. High resolution line lists have energy levels that have been updated either using MARVEL energies (denoted Ma) or energies from some other source with very high accuracy/ low uncertainties. The penultimate column of the line list gives the original calculated energy level; of course if the state energy remains the calculated one (denoted Ca) then these two levels will be the same. The final auxiliary column gives the source of this updated energy level, see Table~\ref{tab:abbr}; the various means of improving on the original energy levels are discussed by McKemmish {\it et al.} \cite{jt948}. 

\begin{table}
    \caption{Extract from the state file for XAB line list for  $^{24}$Ca$^{1}$H \cite{jt858}.}
    \label{t:MgOstates}
    \begin{center}
    \footnotesize
    \tabcolsep=5pt
    \begin{tabular}{rrrrrrrcclrrrr c rrr}
    \hline
    $i$ & \multicolumn{1}{c}{$\tilde{E}$} &  $g$    & $J$  & unc & $\tau$ & $g$
    &\multicolumn{1}{c}{$+/-$} &  \multicolumn{1}{c}{$e/f$} & State & $v$ &
    $\Lambda$& $\Sigma$ & $\Omega$& Abbr &$\tilde{E}_0$ \\
    \hline
    33 & 24428.044063 & 4 & 0.5 & 0.007010 & 1.7190E-07 & 2.002233 & +& e & B'(2SIGMA+)& 3 & 0 & 0.5 & 0.5 & Ma & 24428.027169 \\
    34 & 24986.629171 & 4 & 0.5 & 2.507500 & 5.4398E-08 & -0.000637 & +& e & A(2PI) & 4 & 1 & -0.5 & 0.5 & Ca & 24986.629171 \\
    35 & 25163.713092 & 4 & 0.5 & 0.022010 & 2.0005E-07 & 2.002233 & +& e & B'(2SIGMA+)& 4 & 0 & 0.5 & 0.5 & Ma & 25163.694359 \\
    36 & 25876.111435 & 4 & 0.5 & 0.101011 & 3.1000E-07 & 2.002217 & +& e & B'(2SIGMA+)& 5 & 0 & 0.5 & 0.5 & Ma & 25876.060619 \\
    37 & 26224.114885 & 4 & 0.5 & 3.007500 & 5.8534E-08 & -0.000612 & +& e & A(2PI) & 5 & 1 & -0.5 & 0.5 & Ca & 26224.114885 \\
    \hline
\end{tabular}
\end{center}
\noindent
{
    \flushleft
    $i$: State counting number. \\
    $\tilde{E}$: State energy in \cm. \\
    $g$: State degeneracy. \\
    $J$: Total angular momentum. \\
    unc: Energy uncertainty in \cm. \\
    $\tau$: Lifetime in s. \\
    $g$: Land\'e $g$-factor. \\
    $+/-$: Total parity. \\
    $e/f$: rotationless-parity. \\
    State: state term value using \texttt{PyValem} format \cite{pyvalem,pyvalem2}. \\
    $v$: State vibrational quantum number. \\
    $\Lambda$: Projection of the electronic angular momentum. \\
    $\Sigma$: Projection of the electronic spin. \\
    $\Omega$: $\Omega=\Lambda+\Sigma$, projection of the total angular momentum. \\
    Abbr: Abbreviation giving source of state energy, see Table~\ref{tab:abbr}. \\
    $\tilde{E}_0$: Original (calculated) state energy in \cm. \\
}
\end{table}

\begin{table}
\centering
    \caption{Source type abbreviations used to describe energy levels in hybrid (MARVELized) line lists;
    see  McKemmish {\it et al.} \cite{jt948} for further details.}
    \label{tab:abbr}
    \begin{tabular}{ll}
    \hline
    Abbr & Meaning \\
    \hline
    Ca & Calculated \\
    Ma & \Marvel{} \\
    EH & Effective Hamiltonian \\
    Mo & MOLLIST \\
    HI & HITRAN\\
    PS & Predicted Shift  \\
    IE & Isotopologue Extrapolation \\
    \hline
\end{tabular}

\end{table}

\subsection{API, versioning, master and def files}
\label{SS:API}

The ExoMol database has an application programming interface (API) which allows users to extract data computer to computer without the need
to manually download datasets from the website. Details of how to use the API are given in the ExoMol2020 release \cite{jt810}.

It is ExoMol practice to give each line list a name, denoted \texttt{<DATASET>} below. Alongside the name, each line list
has a version number given by the version date in YYYYMMDD format. MARVELization or other energy level updates to the \texttt{.states} file just lead to 
a new version of the line list. Conversely, a newly calculated line list, and hence a new \texttt{.trans} file, is always
given an new name. Both old versions and old (retired) line lists are retained in the database to allow users to compare with previous results. The most recent version is stored without a version number (date); all previous versions are also retained and have the version in
YYYYMMDD format added to their name.

To facilitate the use of the API and to provide complete metadata, the database contains a \texttt{.master} file, which is located at
\url{www.exomol.com/exomol.all}. The \texttt{.master} file points to a \texttt{.def} for each recommended line list. Note that there are \texttt{.def} and \texttt{\_def.json} files
for each isotopologue for which ExoMol provides a line list. The \texttt{.master} file itself has a version number in YYYYMMDD format;
this date corresponds the latest update of a \texttt{.def}. The version dates of the individual line lists are given in the
 \texttt{.master} file. 

JSON (JavaScript Object Notation) \cite{JSON} is a lightweight human-readable, data-interchange format. ExoMol database newly provides JSON format \texttt{.master} (\texttt{exomol.json}) and definition (\texttt{<ISOTOPOLOGUE>\_\_<DATASET>\_def.json}) files. In the appendix, Listing~\ref{lst:allJSON} gives an extract from the current, JSON format \texttt{.master} file (\texttt{exomol.json}) with keywords defined in Table~\ref{tab:alljsonkeyw} and Listing~\ref{lst:H2OJSON} shows the JSON format definition file (\texttt{\_def.json}) of  the POKAZATEL H$_2$$^{16}$O line list.  Table~\ref{tab:defjsonkeyw} gives the JSON keyword structure
for the \texttt{.def} file.

\begin{table}[H]
\centering
    \small
    \caption{Keywords used in the JSON format \texttt{.master} (\texttt{exomol.json}) file.}
    \label{tab:alljsonkeyw}
    \begin{tabular}{ll}
    \hline
    Keywords & Meaning \\
    \hline
    ID & Always the string \texttt{"EXOMOL.master"} \\
    version & Version number with format YYYYMMDD \\
    num\_molecules & Number of molecules in the database \\
    num\_isotopologues & Number of isotopologues in the database \\
    num\_datasets & Number of datasets in the database \\
    num\_species & Number of species in the database \\
    molecules & List of molecules \\
    $|$\quad formula* & Molecule chemical formula \\
    $|$\quad $|$\quad num\_molecule\_names & Number of molecule names listed \\
    $|$\quad $|$\quad molecule\_names & List  of the molecule names\\
    $|$\quad $|$\quad num\_isotopologues & Number of isotopologues considered \\
    $|$\quad $|$\quad linelist & Line list \\
    $|$\quad $|$\quad $|$\quad inchikey & InChIKey of isotopologue \\
    $|$\quad $|$\quad $|$\quad iso\_slug & Iso-slug \\
    $|$\quad $|$\quad $|$\quad iso\_formula & Formula of isotopologue \\
    $|$\quad $|$\quad $|$\quad dataset\_name & Isotopologue dataset name \\
    $|$\quad $|$\quad $|$\quad version & Version number with format YYYYMMDD \\
    \hline
\end{tabular}
\noindent
{\flushleft \small
*: The keywords with * are variable.
} 
\end{table}

\begin{table}[H]
    \caption{Keywords use in the  JSON format definition (\texttt{<ISOTOPOLOGUE>\_\_<DATASET>\_def.json}) file.}
    \label{tab:defjsonkeyw}
    \resizebox{\textwidth}{!}{
    \begin{tabular}{ll}
    \hline
    Keywords & Meaning \\
    \hline
    isotopologue & Isotopologue section \\
    $|$\quad iso\_formula & Formula of isotopologue \\
    $|$\quad iso\_slug & Iso-slug \\
    $|$\quad inchikey & InChIKey of isotopologue \\
    $|$\quad inchi & InChI of isotopologue \\
    $|$\quad mass\_in\_Da & Mass of isotopologue (sum of atom masses) in  Da \\
    $|$\quad point\_group & Symmetry group \\
    \hline
    atoms & Atoms section \\
    $|$\quad number\_of\_atoms & Number of atoms \\
    $|$\quad element & Element symbol \\
    $|$\quad $|$\quad atom* & In format: Element symbol : Isotope number \\
    \hline
    irreducible\_representations & Irreducible representation section \\
    $|$\quad label* & In format: Irreducible representation label : Nuclear spin degeneracy \\
    \hline
    dataset & Dataset line list files section \\
    $|$\quad name & Isotopologue dataset name \\
    $|$\quad version & Version number with format YYYYMMDD \\
    $|$\quad doi & DOI of dataset paper \\
    $|$\quad max\_temperature & Maximum temperature of linelist \\
    $|$\quad num\_pressure\_broadeners & Number of pressure broadeners available \\
    $|$\quad nxsec\_files & Number of cross section files available \\
    $|$\quad nkcoeff\_files & Number of $k$-coefficient files available \\
    $|$\quad dipole\_available & Dipole availability (true$=$yes, false$=$no, null$=$undefinded) \\
    $|$\quad cooling\_function\_available & Cooling function availability (true$=$yes, false$=$no, null$=$undefinded) \\
    $|$\quad specific\_heat\_available & Specific heat availability (true$=$yes, false$=$no, null$=$undefinded) \\
    $|$\quad continuum & Photo-absorption continuum cross sections availability (true$=$yes, false$=$no, null$=$undefinded) \\
    $|$\quad predis & Photo-dissociation cross sections availability (true$=$yes, false$=$no, null$=$undefinded) \\
    $|$\quad states & States file part \\
    $|$\quad $|$\quad number\_of\_states & Number of states in \texttt{.states} file \\
    $|$\quad $|$\quad max\_energy & Maximum energy (in cm$^{-1}$) in the states file \\
    $|$\quad $|$\quad uncertainty\_description & Uncertainty description \\
    $|$\quad $|$\quad uncertainties\_available & Uncertainty availability (true$=$yes, false$=$no, null$=$undefinded) \\
    $|$\quad $|$\quad lifetime\_available & Lifetime availability (true$=$yes, false$=$no, null$=$undefinded) \\
    $|$\quad $|$\quad lande\_g\_available & Lande g-factor availability (true$=$yes, false$=$no, null$=$undefinded) \\
    $|$\quad $|$\quad num\_quanta & Number of quanta defined \\
    $|$\quad $|$\quad states\_file\_fields & Columns list in the states file \\
    $|$\quad $|$\quad $|$\quad \multirow{2}{*}{name} & Column name in \texttt{.states} file. There are three formats: \\
    $|$\quad $|$\quad $|$\quad & 1. Main column name; 2. Quantum case label : Quantum label; 3. Auxiliary : Auxiliary title \\
    $|$\quad $|$\quad $|$\quad desc & Description of column in the states file \\
    $|$\quad $|$\quad $|$\quad ffmt & Fortran format \\
    $|$\quad $|$\quad $|$\quad cfmt & C format \\
    $|$\quad transitions & Transitions file(s) part \\
    $|$\quad $|$\quad number\_of\_transitions & Total number of transitions in \texttt{.trans} file(s) \\
    $|$\quad $|$\quad number\_of\_transition\_files & Number of \texttt{.trans} file(s) \\
    $|$\quad $|$\quad max\_wavenumber & Maximum wavenumber (in cm$^{-1}$) \\
    \hline
    partition\_function & Partition function file section \\
    $|$\quad max\_partition\_function\_temperature & Maximum temperature of partition function \\
    $|$\quad partition\_function\_step\_size & Step size of temperature \\
    \hline
    broad & Broadening files section \\
    $|$\quad default\_Lorentzian\_half-width & Default value of Lorentzian half-width for all lines (in cm$^{-1}/$bar) \\
    $|$\quad default\_temperature\_exponent & Default value of temperature exponent for all lines \\
    $|$\quad label* & A label for a particular broadener \\
    $|$\quad $|$\quad filename & Filename of particular broadened \\
    $|$\quad $|$\quad max\_J & Maximum $J$ for which pressure broadening parameters provided \\
    $|$\quad $|$\quad Lorentzian\_half\_width & Value of Lorentzian half-width for $J$"$>J_{\textrm{max}}$ \\
    $|$\quad $|$\quad temperature\_exponent & Value of temperature exponent for lines with $J$"$>J_{\textrm{max}}$ \\
    $|$\quad $|$\quad num\_quantum\_number\_sets & Number of defined quantum number sets \\
    $|$\quad $|$\quad quantum\_number\_sets & Defined quantum number sets \\
    $|$\quad $|$\quad $|$\quad code & A code that defines this set of quantum numbers \\
    $|$\quad $|$\quad $|$\quad num\_lines & Number of lines in the broad that contain this code \\
    $|$\quad $|$\quad $|$\quad num\_quantum\_numbers & Number of quantum numbers defined \\
    $|$\quad $|$\quad $|$\quad quantum\_numbers & Defined quantum number(s) \\
    \hline
\end{tabular}
}
\noindent
{\flushleft \small
*: The keywords with * are variable.
}
\end{table}

\subsection{Cross sections and continuum absorption}
\label{ss:xsec}
Table~\ref{tab:cross} gives the general ExoMol format for cross sections  (\texttt{.cross}) files.
Cross sections for line lists are supplied \cite{jt731} to allow users to get an overall picture of the line lists contained in the database. These are provided for different temperatures but at zero pressure (i.e. there is no allowance for pressure broadening) via the ExoMol cross section app allowing to select the temperature, wavenumber range and the resolution (in \cm).

The extension of line lists into the UV means that molecules often have a continuum contribution to their spectrum alongside their characteristic, sharp line absorption.
This temperature-dependent continuum component of the absorption is provided as a set of\texttt{.cont} files given at 100 K intervals from 100~K to the stated $T_{\rm max}$ for the given line list in same format as  the \texttt{.cross} file. The continuum functions were generated neglecting effects due to pressure broadening which are unlikely to be significant for continuum absorptions. The methodology has been introduced in Tennyson {\it et al.} \cite{jt898}.  \texttt{.cont} files are currently available for AlH and AlD \cite{jt922}, and OH \cite{jtOH}. 

\begin{table}[H]
\centering
    \caption{Specification of the \texttt{.cross} cross section and \texttt{.cont}  file
    format}
    \label{tab:cross}
    \begin{tabular}{llll}
    \hline
    Field & Fortran Format & C Format & Description\\
    \hline
    $\tilde{\nu}_i$ & \texttt{F12.6} & \texttt{\%12.6f} & Central bin wavenumber,
    $\mathrm{cm^{-1}}$\\
    $\sigma_i$ & \texttt{ES14.8} & \texttt{\%14.8e} & Absorption cross section,
    $\mathrm{cm^2\,molec^{-1}}$\\
    \hline
    \end{tabular}
    {
    \flushleft
    \noindent
    Fortran format: \texttt{(F12.6,1x,ES14.8)}\\
    }
\end{table}

\subsection{Specific heats and cooling functions}
The specific heat capacity at constant pressure (isobaric) $C_p$ is supplied on a 1~K grid from $T=0$ K to $T_{\rm max}$ in units of 
J~mol$^{-1}$~K$^{-1}$, where available. Otherwise, it can 
be computed from the available partition function by \cite{jt661}
\begin{equation}
    C_p(T) = R \left[\frac{Q''}{Q}-\left(\frac{Q'}{Q}\right)^2 \right] + \frac{3R}{2},
\end{equation}
where the second term is the translational contribution, and 
\begin{equation}
    Q'(T) = T\frac{\mathrm{d} Q}{\mathrm{d} T},
\end{equation}
\begin{equation}
    Q''(T) = T^2\frac{\mathrm{d}^2 Q}{\mathrm{d} T^2} + 2Q',
\end{equation}
and $R$ is the gas constant.    
A complete set of specific heats was generated for the main (parent) isotopologue  for each recommended ExoMol line list
as part of the project to
provide these data in standard (NASA) polynomial form \cite{jt899}.

Cooling functions can be important for astronomical studies \cite{jt489,jt506,jt551,22NgCoPi.exo}
and can be computed from ExoMol line lists.
The temperature-dependent cooling function, $\textsl{W}(T)$ (units erg\,(s\,sr\,molecule)$^{-1}$ $=10^{-7}$ Watts\,(molecule\,sr)$^{-1}$) is the total energy per unit solid angle emitted by a molecule:
\begin{equation} 
    \textsl{W}(T) = \frac{1}{4 \pi \textsl{Q}(T)} \sum_{f,i} A_{fi} h c \tilde{\nu}_{fi} g'_f e^{-c_2 \tilde{E}'_f / T}.
    \label{eq:cooling}
\end{equation}
where $Q$ is the partition function, $A_{fi}$ the relevant Einstein-A coefficient, $ \tilde{\nu}_{fi}$ is the transition wavenumber, $g'_f$ is the degeneracy of the upper state
and $\tilde{E}'_f$ is the energy of the upper state. Finally, $c_2$  is the second
radiation coefficient which allows for an energy level given in terms of wavenumbers.
This can be done using either  \textsc{ExoCross} \cite{jt708} or \textsc{PyExoCross} \cite{jt914}. At present the ExoMol website provides a small number of precomputed
cooling functions.  

\subsection{ExoMolOP: Opacities}
\label{ss:op}

Chubb et al. \cite{jt801} computed
opacity cross sections for all molecules available from the ExoMol database plus the oxygen molecule, and atoms Na and K (using line position data from Kurucz~\cite{11Kurucz.db} and data for the broadening of the strong doublets from Allard {\it et al.} \cite{16AlSpKi.K,19AlSpLe.Na})
for a grid of temperatures and pressures and broadening regime appropriate for hot Jupiters, {\it i.e.} broadening by H$_2$ and He. The data are
provided in formats appropriate for direct use in exoplanetary retrieval and modelling codes ARCiS \cite{ARCiS}, which uses the well established .fits format \cite{fits}, petitRADTRANS \cite{19MoWaBo.petitRADTRANS,23MoNaBl}, NEMESIS \cite{NEMESIS} and Tau-REx \cite{TauRex3}.
The data format and gridding is code dependent with opacities for ARCiS, petitRADTRANS, and NEMESIS  given as $k$-tables at a resolving power of $R$~=~$\frac{\lambda}{\Delta\lambda}$~=~1000 while  cross-sections are provided for Tau-REx with R~=~15\,000. The work of Chubb et al.
has been updated for the newly created ExoMol line lists listed in Table~\ref{tab:exomoldata}. At present these opacities are typically only for the main (parent) isotopologue for each molecule (with some exceptions, for example CO is available with isotopologues mixed at terrestrial natural abundances).

The plan for the future will be to provide a broader range of opacities which (a) allow for mixtures of isotopologues and (b) using our newly derived broadening parameters (see below) to provide opacities for a variety of atmospheric compositions.

\section{Collisional line broadening}\label{s:broad}


The generation of opacities requires knowledge of line shape parameters, with collisional line broadening being a key process \cite{jt773} in determining the frequency distribution of a line's intensity. 
In the course of previous works in the group, an ExoMol diet was introduced \cite{jt684} which provided parameters characterising the collisional line broadening required for the Voigt profile. The reference collisional line width $\gamma_0$ (in cm$^{-1}$atm$^{-1}$) and its temperature exponent $n$ are assumed to obey the following power law:
\begin{equation}
\label{gammaPL}
\gamma(T,P)=\gamma_0(T_{\rm ref}, P_{\rm ref}) \left(\frac{T_{\rm ref}}{T}\right)^n \frac{P}{P_{\rm ref}},
\end{equation}
where $T_{\rm ref}$, $P_{\rm ref}$ are some reference temperature and pressure, which in ExoMol are chosen as 1~atm and $296$~K, respectively. 
The ExoMol diet tabulates  Lorentzian half-width, $\gamma_0$, and its temperature dependence, $n$, for each perturbing species, and the final half-width is calculated as a sum of contributions from all the perturbers. The ExoMol diet is organised in the form of `recipes', tabulating  the  dependence of $\gamma_0$ and $n$  on the rotational and vibrational quantum numbers.

The pressure broadening \texttt{.broad} file structures used by the ExoMol diet are illustrated in Table~\ref{t:broad:diet}. 
For each radiator-perturber pair, the diet files contain one or more recipes indicated by a unique name in the first column (\texttt{a0}, \texttt{a1}, etc.). This name is followed by collisional broadening parameters, for instance $\gamma_0(T_\text{ref},P_\text{ref})$ and $n$ from Eq.~\eqref{gammaPL}, and by a set of quantum numbers used in the recipe's parametrization scheme. For example, $J''$ is used in \texttt{a0}, while \texttt{a1} additionally uses $K''$. Note that a file might have a varying number of columns if different recipes require it.
The naming convention for all pressure broadening files is to give broadening data for the main isotopologue, however pressure broadening data are not isotope specific in the current iteration of the database. If a user wishes to include pressure broadening for minor isotopologues, then they should use the data provided for the major isotopologue. 

\begin{table}[H]
\caption{Examples of the .broad ExoMol diet files for the `a0', `a1' and `m0' recipes. }
\label{t:broad:diet} \footnotesize
\begin{center}
\begin{tabular}{crrrrr}
\hline
\hline
Label & $\gamma$ & $n$ & $J''$/$m$ & $K''$  \\
\hline
\multicolumn{4}{c}{\texttt{24Mg-16O\_\_O2.broad}}\\
a0 & 0.043 & 0.500 &      0\\
... & & & \\
\multicolumn{4}{c}{\texttt{1H2-16O\_\_H2.broad}}\\
a1 &0.0916& 0.790 &      0 &      1\\
a1 &0.0852& 0.608 &      1 &      2\\
a1 &0.0764& 0.541 &      2 &      3\\
a1 &0.0699& 0.502 &      3 &      4\\
a1 &0.0635& 0.458 &      4 &      5\\
... & & & \\
\multicolumn{4}{c}{\texttt{27Al-1H\_\_H2.broad}}\\
m0& 0.1554& 0.6406 &     1\\
m0& 0.1489& 0.6235 &     2\\
m0& 0.1450& 0.6128 &     3\\
m0& 0.1410& 0.6002 &     4\\
... & & & \\
\hline
\end{tabular}
\end{center}
\end{table}

Table~\ref{t:broad:diets} list the current menu of ExoMol diets.  
As a new diet option,  the spectroscopic index $|m|$ has been introduced  to better describe the rotational dependence of collisional parameters $\gamma_0$ and $n$,  where  $m$ is defined as 
$m = -J^{''} $ for P-branch transitions,
$m = J^{''}+1 $ for R-branch, and
$m = J^{''} $ for Q-branch (i.e. the same as done in HITRAN). The ``m0'' option (see Table~\ref{t:broad:diets})  has been used to the define $\gamma_{\rm L}$ for AlH as part of the recent AloHa line list  \cite{jt922} (labeled in the diet \texttt{.broad} file as \texttt{m0}); the ``m1'' recipe was introduced  to describe the dependence on the  lower state $K''$ quantum number for symmetric top molecules  in addition to the dependence on $|m|$ \cite{jt952}.

\begin{table}[H]
    \centering
    \caption{
        Parameters defined in the ExoMol broadening diets.}
        \label{t:broad:diets} 
    \begin{tabular}{cll}
    \hline
    diet code &  quantum numbers & systems \\
    \hline
    \texttt{a0} & 
        $J''$
         & general use (e.g. \cite{jt684}) \\
    \texttt{a1} &
        $J''$, $K''$ & general use (e.g. \cite{jt684}) \\
    \texttt{m0} &
        $\mid \! m \! \mid$ & general use (e.g.  \cite{jt922}) \\
    \texttt{m1} &
         $\mid \! m \! \mid$, $K''$ & general use (e.g. \cite{jt952}) \\
    \texttt{a5} & 
        $J'$, $K'_a$, $K'_c$, $J''$, $K''_a$, $K''_c$
         & asymmetric tops ( e.g. SO$_2$ \cite{jt684}) \\
    \hline
    \end{tabular}
\end{table}

 The ExoMol database contains empirical broadening data for seven species: \ce{H2O}, \ce{NH3}, \ce{SO2}, \ce{CH4}, \ce{PH3}, \ce{HCN}, and \ce{H2CO} broadened by \ce{H2} and \ce{He}. In addition, for most of the other ExoMol molecules, especially exotic ones, the ExoMol diet now provides theoretical line broadening parameters computed using at least one of three methods,  simple semi-classical calculations,  explicit semi-classical Modified Complex Robert-Bonamy (MCRB) approach and Machine Learning  described below. 


The MCRB semi-classical approach \cite{07MaTiBo2.broad} is capable of producing broadening parameters accurate to within experimental uncertainty 
if an accurate interaction potential between the radiating and perturbing molecule is available. If only rough model potential data are available, fitting of the interaction potential to  experimental line broadening coefficients can be used to improve  the description for a range of transitions and temperatures. Calculations of vibrational contributions to broadening coefficients additionally require the knowledge of ro-vibrational state-dependent molecular quantities for the radiating molecule (dipole moment, polarizability). Broadening calculations have been done for \ce{AlH} broadened by \ce{H2}, \ce{He}, \ce{N2}, and \ce{AlH} \cite{jt922}, and work on \ce{C2H2} broadened by \ce{H2}, \ce{He}, \ce{N2}, \ce{C2H2}, and \ce{CO2} is currently under way \cite{jtHCCHgamma}. In these cases, the vibrational contributions have been found to be negligible, so the respective \texttt{.broad} files only contain the dependence on the rotational quantum number $J$. Future work includes producing $J$ ($m$)-dependent broadening parameters for a number of diatomic molecules from the ExoMol database broadened by various perturbers using model interaction potentials, where the calculations are not too demanding.

A novel  Machine Learning (ML) method  has been  developed \cite{jt919}  to predict the line  broadening parameters  $\gamma_0$. This ML approach is suitable for the cheap mass production of pressure broadening parameters for any active neutral molecule; the ML code and associated data available on the ExoMol Zenodo area \cite{24Guest.broad}. The algorithm considers the rotational dependence of the broadening but ignores any dependence on vibrational and/or electronic states. 

Using this ML code, $J$-dependent air-broadening parameters   $\gamma_{\rm air}$ have been produced for all molecules in the ExoMol database, suitable for room temperature and pressure environments, where air is the primary source of broadening.The empirical $\gamma_0^{\rm air}$ values for training were taken from the HITRAN database  \cite{jt857}.  These data are available on the ExoMol website, where the temperature exponent value of $n=0.5$ is assumed.  Further work using machine learning to predict $\gamma_0$ for other perturbing species is ongoing. The molecules for which $J$-dependent air-broadening data are available is summarised in Table~\ref{t:broad:data_ML}. High quality data have simply been copied from the HITRAN database \cite{jt857} while other values are machine learned predictions \cite{jt919}.

\begin{table}
    \centering
    \caption{\label{t:broad:data_ML} Availability of new air broadening data, $\gamma_{\rm air}$. The molecules with ML data were generated with \cite{jt919}. Other air broadening data are taken from the HITRAN database \cite{jt857}.} 
    \begin{tabular}{lllll|ll}
    \hline
    \multicolumn{5}{c}{ML data} & \multicolumn{2}{c}{HITRAN data} \\
    \hline
    \ce{MgH} & \ce{NaH} & \ce{NiH} & \ce{AlH} & \ce{CrH} & \ce{HCl} & \ce{CH3Cl}\\
    \ce{CaH} & \ce{BeH} & \ce{TiH} & \ce{FeH} & \ce{LiH} & \ce{HF} & \ce{C2H4}\\
    \ce{ScH} & \ce{NH} & \ce{CH} & \ce{OH} & \ce{SiH} & \ce{HBr} & \ce{C2H2}\\
    \ce{SH} & \ce{PH} & \ce{VO} & \ce{ALO} & \ce{YO} & \ce{CO} & \ce{PH3}\\
    \ce{MgO} & \ce{TiO} & \ce{SiO} & \ce{CaO} & \ce{NaO} & \ce{NO} & \ce{H2O}\\
    \ce{LaO} & \ce{ZrO} & \ce{SO} & \ce{PO} & \ce{PN} & \ce{O2} & \ce{CO2}\\
    \ce{KCl} & \ce{NaCl} & \ce{LiCl} & \ce{CN} & \ce{C2} & \ce{CS} & \ce{SO2}\\
    \ce{H2} & \ce{CP} & \ce{PS} & \ce{NS} & \ce{SiS} & \ce{N2} & \ce{HCN}\\
    \ce{NaF} & \ce{AlCl} & \ce{AlF} & \ce{KF} & \ce{LiF} & \ce{CH4} & \ce{N2O}\\
    \ce{CaF} & \ce{MgF} & \ce{SiN} & \ce{H2O2} & \ce{SO3} & \ce{NH3} & \ce{H2S}\\
    \ce{SiH4} & \ce{AsH3} & \ce{PF3} & \ce{CH3} & \ce{P2H2} & \ce{HNO3} & \ce{OCS}\\
    \ce{H2CS} & \ce{CaOH} & \ce{KOH} & \ce{NaOH} & \ce{SiH2} & \ce{H2CO} & \\
    \ce{SiO2} & \ce{LiOH} & & & & \ce{CH3F} & \\
    \hline
    \end{tabular}
\end{table}


The need for at least approximate values of pressure-broadened rovibrational line widths for ``exotic'' molecular pairs relevant to exoplanetary atmospheres initiated a theoretical study \cite{jt872} giving access to simple semi-classical (rotationally-independent) estimates from the knowledge of molecular masses and kinetic diameters. Being valid for neutral active molecules but also for spectroscopically active molecular ions, this method allowed population of the ExoMol database with estimates of pressure-broadening coefficients for perturbation by H$_2$, O$_2$, N$_2$, He, Ar, CO, CS, NO, CO$_2$, H$_2$O, CH$_4$, NH$_3$ and self-perturbation for 45 active molecules (main species) \cite{jtSC} in ExoMol, see Table~\ref{t:broad:data0}; work is in progress on producing similar data for their most abundant isotopologues. Available on the ExoMol website are the reference-temperature (296 K) predictions issued from data considered as the most reliable/recent kinetic diameter values. Results for  CH$_2$, FeO, PO$_2$, SH$_3$, SiC, SiH$_3$ and TiF are not currently available on the ExoMol website, as there are no corresponding line lists available for these molecules. Results for temperatures above 296~K are easily obtained by a simple scaling factor $\sqrt{296/T}$ but the temperature considered should not be too low as classical-path approximation has to remain valid (i.e. the kinetic energy of the relative molecular motion should be higher than the isotropic intermolecular potential depth), see the discussion by Buldyreva {\it et al.} \cite{jtSC}.

\begin{table}
    \centering
    \caption{\label{t:broad:data0} New (rotationally-independent) line broadening data $\gamma_0$ in the ExoMol database generated using the semi-classical method by Buldyreva {\it et al.}  \cite{jt872}.}
    \begin{tabular}{llll|l}
    \hline
    \multicolumn{4}{c}{Active molecules} & Perturbers \\
    \hline
AlCl     & CrH      & NaCl     & SH       & self     \\
AlH      & CS       & NaF      & SiH      & Ar       \\
AlO      & KCl      & NaH      & SiH$_2$  & CH$_4$   \\
AsH$_3$  & KF       & NaOH     & SiO      & CO       \\
BeH      & KOH      & NH       & SiO$_2$  & CO$_2$   \\
C$_3$    & LiCl     & NS       & SiS      & H$_2$    \\
CaF      & LiF      & PF$_3$   & TiH      & H$_2$O   \\
CaH      & LiH      & PH       & TiO      & He       \\
CaO      & LiH$^+$  & PN       & VO       & N$_2$    \\
CaOH     & MgF      & PO       &          & NH$_3$   \\
CH$_3$   & MgH      & PS       &          & NO       \\
CP       & MgO      & ScH      &          & O$_2$    \\
    \hline
    \end{tabular}
\end{table}

Pressure-induced line-shape parameters have also been addressed for the case of vibronic transitions connecting low-lying vibrational levels in the ground and an excited (bound) electronic states \cite{jt907}. In this case the collisional (pressure) broadening is still not dominated by predissociative effects. Pre-computed potential-energy surfaces for interactions of some representative molecular pairs enabled both line-width and line-shift calculations comparing favourably with available measurements from room temperature to 2800~K \cite{jt907}. Further attempts at improvements are in progress, in order to establish a reliable computational scheme for providing line widths and shifts for rovibronic transitions of active molecular species listed in ExoMol.

In instances where a new set of broadening data is introduced for a system with already available data, a version number is added to the filename of the older data set in order to distinguish multiple versions of the same system. The version number is a timestamp indicating the date the dataset was created (uploaded to the data base) in the YYYYMMDD format as described in Sect.~\ref{SS:API}. The most recent or recommended  version does not contain a version number. 
Although, at present, there is only one molecule, AlH, with broadening parameters computed by more than one method, we expect this number to grow. Apart from the experimental sources as e.g. provided by HITRAN, line broadening parameters have being provided by different computational methods such as ML \cite{jt919}, MCRB \cite{07MaTiBo2.broad},  simple semi-classical  \cite{jt872,jtSC} as well phase shift \cite{jt907} approaches.


\section{Predissociative line Broadening}\label{s:prediss}

As introduced in Sect.~\ref{ss:data}, the definition of the lifetimes in ExoMol has been extended to include the predissociative lifetimes as part of the \texttt{.states} file provision. The practical purpose of this extension is to be able to model the predissociative broadening together with the collisional and Doppler broadening. The lifetime broadening is described by  Lorentzian line shape like pressure broadening with  $\gamma_\tau$, the half-width in cm$^{-1}$,  given by  
\begin{equation}
\gamma_\tau = \frac{\hbar}{2\tau h c}, 
\end{equation} 
where $\tau$ is the half-harmonic mean of the radiative $\tau_{\rm r}$ and predissociative $\tau_{p}$ lifetimes via
\begin{equation} \label{eq:tau_tot}
    \tau = \frac{\tau_{\rm r} \tau_{\rm p}}{\tau_{\rm r}+\tau_{\rm p}}. 
\end{equation}
The half-width $\gamma_\tau$ can  be added to the pressure-broadening half-width, $\gamma$ in Eq.~\eqref{gammaPL}, to give the total Lorentzian component of the line profile. 

Predissociation due to rotational barrier effects has been studied for AlH, see sec. \ref{subsubsec:AloHa}. Predissociative line broadening due to spin-orbit interactions with the continuum has also been studied for OH \cite{jt933}, see \ref{subsection:hydroxyl},  using a modified stabilization method \cite{70HaTaxx,82BaSi,93MaRaTa,jtpred}. Predissociation lifetimes will also soon be released for SH and SD using the spectroscopic model from current linelist \cite{jt725, 19GoYuTe} with the addition of the dissociative states, 1~$^2 \Sigma^-$, 1~$^4 \Sigma^-$, and 1~$^4 \Pi$.


\section{Photodissociation and continuum absorption}
\label{s:photo}

The extension of the database to consider photodissociation means we define a new file type \texttt{.photo}, which provides temperature-dependent
photodissociation cross sections. The format of the photodissociation cross section has been introduced in Tennyson {\it et al.} \cite{jt898} and specified in Table~\ref{tab:photo}. In contrast to other data in the database, \texttt{.photo} files give results as a function of wavelength in nm. The project is only just
starting to generate photodissociation cross sections, and so far, calculated cross sections are available for HF and HCl \cite{jt865}. A major expansion of this part of the database is planned.

The previous section in the database/on the website called photo-absorption continuum cross sections, which contained measured UV photoabsorption cross sections,
had been moved into the photodissociation area. These data are presented in the same format as the computed photodissociation
cross sections, but the temperature, pressure and broadening parameters are determined by the experimental conditions used and
generally do not form a regular grid. Measured cross sections are available for \ce{H2O}, \ce{CO2}, \ce{SO2}, \ce{NH3}, \ce{H2CO}, \ce{C2H4} and \ce{CO}. These cross sections probably represent a good approximation to the photodissocition cross section for the given species \cite{18VeBeFa.CO2}. Table 4 of the ExoMol2020 release \cite{jt810} gives more details on these cross sections provided which are  
due to Venot {\it et al.} \cite{18VeBeFa.CO2} and Fateev and co-workers (private communication, 2020).

\begin{table}[H]
  \centering
    \caption{Specification of the  wavelength-dependent \texttt{.photo} cross section file
    format}\label{tab:photo}
    \begin{tabular}{llll}
    \hline
    Field & Fortran Format & C Format & Description\\
    \hline
    $\lambda_i$ & \texttt{F6.2} & \texttt{\%6.2f} & Central bin length, nm\\
    $\sigma_i$ & \texttt{ES14.8} & \texttt{\%14.8e} & Photodissociation cross section, $\mathrm{cm^2\,molec^{-1}}$\\
    \hline
    \end{tabular}
    {
    \flushleft
    \noindent
    Fortran format: \texttt{(F6.2,1x,ES14.8)}\\
    }
\end{table}

\section{Post-processing}\label{s:exocross}

ExoMol provides post-processing capabilities through the Fortran program \textsc{ExoCross} (GitHub: \url{https://github.com/ExoMol/ExoCross}) \cite{jt708} and the Python program \textsc{PyExoCross}  \cite{jt914} (GitHub: \url{https://github.com/ExoMol/PyExoCross}). There are example jobs for both \textsc{ExoCross} and \textsc{PyExoCross} on their GitHub pages.

\textsc{ExoCross} has many functions such as generating pressure and temperature dependent cross sections, partition functions, specific heat,  state-resolved radiative lifetimes, non-local thermodynamic equilibrium (non-LTE) spectra, electric dipole, electric quadrupole and magnetic dipole spectra. \textsc{ExoCross} can read data in both ExoMol and HITRAN \cite{jt557} formats and output them in these formats as well as SPECTRA (\url{http://spectra.iao.ru/}) and Phoenix formats \cite{PHOENIX}. Should data be required in HITRAN format, it is strongly recommended that the data are downloaded to a local computer in the much more compact ExoMol format and then processed using \textsc{ExoCross}.

As a Python adaptation of the post-processing program \textsc{ExoCross}, \textsc{PyExoCross} supports importing and exporting line lists in the ExoMol and HITRAN  format \cite{jt763} and can convert data format between ExoMol and HITRAN formats. \textsc{PyExoCross} provides functions for calculating partition functions, specific heats, radiative lifetimes, cooling functions, oscillator strengths, LTE and non-LTE absorption and emission stick spectra and cross sections. 
Users can extract low uncertainties, high intensities and specified quantum number labels and values by using uncertainty, threshold and quantum number filters when calculating stick spectra and cross sections.
\textsc{PyExoCross} can also help users automatically download line lists files from the ExoMol database in batches. 
The user instruction manual is available on \url{https://pyexocross.readthedocs.io/}.

\section{New web services}\label{ss:newweb}

\subsection{ExoMolHR}
\label{ss:HR}
ExoMolHR (ExoMol High-Resolution) \cite{jtHR} is a new high-resolution molecular spectroscopic database based on the ExoMol line list database; it is  available at \href{www.exomol.com/exomolhr}{www.exomol.com/exomolhr}.
The ExoMolHR database is focused on providing high accuracy line positions for high-resolution studies such as line identification or simulations of high-resolution spectra. Initially reduced line lists are constructed by scraping the ExoMol database for energy levels (and hence transitions) which are determined to high accuracy (uncertainty $\leq 0.01$~cm$^{-1}$). After this all transitions
which can be predicted at high resolution ($R>100\,000$) are stored; these transitions provide the core of the database.
For each transition, ExoMolHR provides the frequency, uncertainty, Einstein $A$-coefficient, intensity (at a user specified temperature), lower energy, total level degeneracy, angular momentum and quantum numbers for upper and lower states. Users can download datasets using the application programming interface (API) directly with the intensities computed at a reference temperature $T_{\textrm{ref}}=296$ K. Alternatively, they can use the interactive
web interface where one can set parameters to filter the wavenumber range, extract low uncertainties and lines with strong intensities, and then store the results generated for downloading. The ExoMolHR database currently contains data for 55 isotopologues from 32 molecules, see Table~\ref{tab:HRlist}; this number is growing quite rapidly as more MARVEL studies are completed. 

\begin{table}
\footnotesize
\centering
\setlength{\tabcolsep}{4.2mm}{
\caption{Contents of the ExoMolHR database  \cite{jtHR}; lines are extracted from the  ExoMol database on the basis that  they have  low uncertainties. 
}
\label{tab:HRlist}
\begin{tabular}{rllrrrrrrrrr}
\hline
ID & Molecule & Isotopologue & Dataset & $N_{\textrm {states}}$\  & $N_{\textrm {files}}$ & $N_{\textrm {trans}}$\ \ \ \ \ \  & $N_{\textrm {HRstates}}$ & $N_{\textrm {HRlines}}$ \\
\hline
1  & AlCl       & $^{27}$Al$^{35}$Cl       & YNAT      & 65869    & 1   & 4722048      & 41    & 101       \\
2  & AlCl       & $^{27}$Al$^{37}$Cl       & YNAT      & 67507    & 1   & 5748704      & 41    & 121       \\
3  & AlH        & $^{27}$Al$^1$H           & AloHa     & 1364     & 1   & 29725        & 135   & 692       \\
4  & AlO        & $^{26}$Al$^{16}$O        & ATP       & 93350    & 1   & 4866540      & 4783  & 143197    \\
5  & AlO        & $^{27}$Al$^{16}$O        & ATP       & 94862    & 1   & 4945580      & 4980  & 149577    \\
6  & AlO        & $^{27}$Al$^{17}$O        & ATP       & 96350    & 1   & 5148996      & 4787  & 142905    \\
7  & AlO        & $^{27}$Al$^{18}$O        & ATP       & 98269    & 1   & 5365592      & 4799  & 142976    \\
8  & C$_2$      & $^{12}$C$_2$             & 8states   & 44189    & 1   & 6080920      & 8376  & 445682    \\
9 & C$_2$H$_2$ & $^{12}$C$_2^1$H$_2$       & aCeTY     & 5160803  & 100 & 4347381911   & 8898  & 473850    \\
10 & CH$_4$     & $^{12}$C$^1$H$_4$        & MM        & 9155208  & 121 & 50395644806  & 21021 & 7649736   \\
11 & CN         & $^{12}$C$^{14}$N         & Trihybrid & 28004    & 1   & 2285103      & 4833  & 244808    \\
12 & CO$_2$     & $^{12}$C$^{16}$O$_2$     & UCL-4000  & 3480477  & 20  & 2557549946   & 18881 & 2600218   \\
13 & CaH        & $^{40}$Ca$^1$H           & XAB       & 6825     & 1   & 293151       & 1165  & 12341     \\
14 & CaOH       & $^{40}$Ca$^{16}$O$^1$H   & OYT6      & 3187522  & 18  & 23384729495  & 1424  & 12984     \\
15 & H$_2$CO    & $^1$H$_2^{12}$C$^{16}$O  & AYTY      & 10297025 & 100 & 12688112669  & 4813  & 317729    \\
16 & H$_2$CS    & $^1$H$_2^{12}$C$^{32}$S  & MOTY      & 52292454 & 8   & 43561116660  & 3625  & 72218     \\
17 & H$_2$O     & $^1$H$_2^{16}$O          & POKAZATEL & 810269   & 412 & 5745071340   & 14395 & 3520554   \\
18 & H$_2$S     & $^1$H$_2^{32}$S          & AYT2      & 220631   & 35  & 115032941    & 2061  & 63719     \\
19 & H$_3$O$^+$ & $^1$H$_3^{16}$O$^+$      & eXeL      & 1173114  & 100 & 2089331073   & 232   & 1785      \\
20 & H$_3^+$    & $^1$H$_2^2$H$^+$         & ST        & 33330    & 1   & 22164810     & 109   & 646       \\
21 & H$_3^+$    & $^1$H$_3^+$              & MiZATeP   & 158721   & 1   & 127542657    & 994   & 13606     \\
22 & H$_3^+$    & $^2$H$_2^1$H$^+$         & MiZo      & 369500   & 32  & 2290235000   & 115   & 683       \\
23 & H$_3^+$    & $^2$H$_3^+$              & MiZo      & 37410    & 21  & 36078183     & 115   & 225       \\
24 & LiOH       & $^6$Li$^{16}$O$^1$H      & OYT7      & 192412   & 5   & 294573305    & 255   & 840       \\
25 & LiOH       & $^7$Li$^{16}$O$^1$H      & OYT7      & 203762   & 5   & 331274717    & 240   & 749       \\
26 & MgH        & $^{24}$Mg$^1$H           & XAB       & 3148     & 1   & 88575        & 237   & 2462      \\
27 & MgH        & $^{25}$Mg$^1$H           & XAB       & 3156     & 1   & 88776        & 548   & 5850      \\
28 & MgH        & $^{26}$Mg$^1$H           & XAB       & 3160     & 1   & 88891        & 537   & 5339      \\
29 & N$_2$O     & $^{14}$N$_2^{16}$O       & TYM       & 1759068  & 21  & 1360351722   & 17018 & 3459640   \\
30 & NH         & $^{14}$N$^1$H            & kNigHt    & 4076     & 1   & 327014       & 1030  & 26131     \\
31 & NH         & $^{14}$N$^2$H            & kNigHt    & 7406     & 1   & 778105       & 118   & 943       \\
32 & NH         & $^{15}$N$^1$H            & kNigHt    & 4089     & 1   & 327877       & 118   & 943       \\
33 & NH         & $^{15}$N$^2$H            & kNigHt    & 7465     & 1   & 785940       & 118   & 943       \\
34 & NH$_3$     & $^{14}$N$^1$H$_3$        & CoYuTe    & 5095730  & 200 & 16941637250  & 4720  & 412149    \\
35 & NO         & $^{14}$N$^{16}$O         & XABC      & 30811    & 1   & 4596666      & 3044  & 106711    \\
36 & OCS        & $^{16}$O$^{12}$C$^{32}S$ & OYT8      & 2399110  & 10  & 2527364150   & 5198  & 279273    \\
37 & PN         & $^{31}$P$^{14}$N         & PaiN      & 30327    & 1   & 1333445      & 32    & 44        \\
38 & SO         & $^{32}$S$^{16}$O         & SOLIS     & 84114    & 1   & 7086100      & 536   & 2501      \\
39 & SO$_2$     & $^{32}$S$^{16}$O$_2$     & ExoAmes   & 3270271  & 80  & 1300000000   & 14924 & 1504495   \\
40 & SiN        & $^{28}$Si$^{14}$N        & SiNfull   & 131936   & 1   & 43646806     & 99    & 670       \\
41 & SiN        & $^{28}$Si$^{15}$N        & SiNfull   & 133460   & 1   & 44816182     & 56    & 464       \\
42 & SiN        & $^{29}$Si$^{14}$N        & SiNfull   & 132335   & 1   & 43946969     & 56    & 464       \\
43 & SiN        & $^{30}$Si$^{14}$N        & SiNfull   & 132706   & 1   & 44223730     & 56    & 464       \\
44 & SiO        & $^{28}$Si$^{16}$O        & SiOUVenIR & 174250   & 1   & 91395763     & 911   & 8729      \\
45 & TiO        & $^{48}$Ti$^{16}$O        & Toto      & 301245   & 1   & 58983952     & 8725  & 499775    \\
46 & VO         & $^{51}$V$^{16}$O         & HyVO      & 3410598  & 90  & 58904173243  & 7043  & 635722    \\
47 & YO         & $^{89}$Y$^{16}$O         & BRYTS     & 173621   & 1   & 60678140     & 28    & 25        \\
48 & YO         & $^{89}$Y$^{17}$O         & BRYTS     & 182598   & 1   & 62448157     & 28    & 25        \\
49 & YO         & $^{89}$Y$^{18}$O         & BRYTS     & 182547   & 1   & 64164605     & 28    & 25        \\
50 & ZrO        & $^{90}$Zr$^{16}$O        & ZorrO     & 227118   & 1   & 47662773     & 5313  & 145317    \\
51 & ZrO        & $^{91}$Zr$^{16}$O        & ZorrO     & 227118   & 1   & 47748501     & 1058  & 5164      \\
52 & ZrO        & $^{92}$Zr$^{16}$O        & ZorrO     & 227124   & 1   & 47830250     & 1058  & 5164      \\
53 & ZrO        & $^{93}$Zr$^{16}$O        & ZorrO     & 227126   & 1   & 47928979     & 1058  & 5164      \\
54 & ZrO        & $^{94}$Zr$^{16}$O        & ZorrO     & 227128   & 1   & 47994352     & 1058  & 5164      \\
55 & ZrO        & $^{96}$Zr$^{16}$O        & ZorrO     & 227134   & 1   & 48136388     & 1058  & 5164      \\
\hline
\end{tabular}
\noindent
{\flushleft \footnotesize
\item $N_{\textrm {states}}$: Number of states in states file; \\
\item $N_{\textrm {files}}$: Number of transitions files; \\
\item $N_{\textrm {trans}}$: Number of transitions in transitions file(s); \\
\item $N_{\textrm {HRstates}}$: Number of states with uncertainties $\le 0.01$ cm$^{-1}$; \\
\item  $N_{\textrm {HRlines}}$: Number of lines in ExoMolHR dataset with $R>100\,000$.\\
}
}
\end{table}

\subsection{LiDB}
\label{ss:LiDB}
LiDB (Lifetimes Database) is a newly developed database of molecular vibrational and vibronic state radiative lifetimes~\cite{jt904}, created to enable radiative effects to be properly captured in low-temperature plasma models. The main data output of LiDB is radiative lifetimes at vibrational and electronic state resolution. Partial lifetimes, which give information on the dominant decay channels in a molecule, are also provided. Datasets for 36 molecules are available, produced from the respective molecular line list in the ExoMol database. LiDB is freely available and hosted at \href{www.exomol.com/lidb}{www.exomol.com/lidb}. Users can dynamically view molecular datasets or use a specially-designed API to perform data requests. LiDB is linked directly to the Quantemol DataBase (QDB) \cite{jt861} which aims to provide comprehensive datasets for plasma modelling. LiDB will expand in the future with the addition of more molecules, important isotopologues, and neutral and singly-charged atomic species.

\section{Future development}

The ExoMol project will continue not only to provide high quality data on more molecules, but the type of data available will be expanded to maximise the scientific richness that can be extracted from observations possible with new and upcoming telescopes.
We are also open to hosting high quality, relevant datasets (both computed and measured) obtained by other groups.

The start in computing photodissociation cross sections and rates, as well as the treatment of rovibronic spectra signals our intention to extend the database to cover processes which occur
at ultraviolet (UV) wavelengths. This will mean developing methods of performing calculations involving many electronic states for molecules larger than diatomics. 

The calculations on the Renner-Teller system of CaOH \cite{jt878} were performed using the variational rovibronic nuclear motion code EVEREST \cite{EVEREST}. We are currently using EVEREST to explore the calculation of photodissociation
spectra of triatomic systems using HCN as a test case \cite{jtHCN}. The plan will be to start to provide extended coverage of temperature-dependent photoabsorption cross sections.
For photodissociation we have already identified a number of studies on diatomics which provide temperature-dependent cross sections  \cite{03WeScSt.MgH,11MiGaSt.HeH+,13ElSt.CN,16McShMc.SH+,18PaCiSt.CS,21QiBaLi.MgO, 21QiBaLi.AlH, 22QiBaLi.AlCl,22QiBaLi.AlF,24BaYaQi.NaO,23QiPeBa.O2}
which we hope to add to the database.

The methodology for generating atomic as opposed to molecular data is actually very different and thus done by different research groups with information dispersed. However, astronomically, both species are often present together; for example, there is now a significant number of neutral and singly charged atoms that have been observed in exoplanetary atmospheres \cite{18SpSiEv.exo,20HoSePi.exo,21CoYaRe.exo}.  To support exoplanet studies, therefore, users have requested that atomic data also be provided in the same ExoMol format to facilitate more convenient analysis. We have started to address this need with the inclusion of the opacities for Na and K in a H$_2$/He atmosphere, see Sect.~\ref{s:atoms}. Further expansions to include atomic data are being actively pursued with  NIST \cite{NIST-www,NIST} and/or the Kurucz database \cite{11Kurucz.db} as possible sources.

As discussed in Sect.~\ref{s:broad} there is a real need for an extended set of broadening parameters. Clearly this is a major focus of on-going work. At present ExoMol opacities and others are largely all focused on hot Jupiter planets. An extended set of broadening parameters would allow the generation of opacities for other atmospheres such as lava planets and rocky planets. What we do have as a result of our machine learning study is a full set of air-broadening parameters. This means we are already in a position to generate a full set of opacities for a planet whose atmosphere is Earth-like, ie close to 80\%\ N$_2$ and 20\%\ O$_2$. The use of different methodologies to generate broadening parameters, as discussed in Sect.~\ref{s:broad}, means that it is possible to have more than one file containing broadening parameters for a given perturber and molecule combination. Generalization of the ExoMol data structure to allow for this possibility will form part of an upcoming paper on the revised ExoMol broadening diet.

Currently, the ExoMol molecular opacities are generated assuming the LTE conditions, where the temperature dependent populations of the molecular states obey the Boltzmann distribution. The non-LTE processes assume a departure from LTE thus making the ExoMol opacities non-applicable for modelling, e.g., the solution of the radiative transfer equation at non-LTE conditions. Some work on using our data for non-LTE studies has already been performed \cite{jt903}. However, a more general solution will require a new cross section format capable of  preserving the information on the (rotation-)vibrational state that is therefore needed to allow for the solution of the statistical equilibrium equation. This will be addressed in the near future. 

Finally but importantly, a major driver in our choice of species and processes to study is requests from interested users.
We are always happy to receive such requests although we do not guarantee rapid delivery of results.

\section{Conclusions}

Since the ExoMol2020 data release \cite{jt810}, 27 new molecules have been added to 
the database. For many others the coverage for the line lists has been extended into
the ultra violet and/or the energy levels, and hence transition wavenumbers, have been
made significantly more accurate. Indeed,  
ExoMol-style line lists have proven extremely effective in providing very high quality data for astronomers, especially when supported by \Marvel{} studies of available experimental data.

The period since the last release of the ExoMol database in 2020 has seen the launch of JWST, the most important space observatory in decades, and its impact on astronomy research particularly in exoplanet atmospheric characterisation has been immediate. The ExoMol line list data have proven extremely powerful for modelling JWST observations when the molecule was predicted, with no identified shortcomings in the ExoMol databases' line list modelling of known molecular spectral bands at different temperatures. Yet spectral bands of unknown molecular origin are being identified regularly. Each of these unidentified spectral bands promises to shed light on physics, chemistry, geology (or maybe even biology) that was unsuspected in planetary models for exoplanetary systems. The set of molecules for which spectral data are required is clearly larger than already available in ExoMol, yet shortlisting the set of likely molecular candidates is not straightforward \citep{23ZaPeMc.exo}, especially for unusual non-terrestrial exoplanets. Computationally generated approximate spectral data for a very large number of potential atmospheric molecules \cite{19SiPeSe.exo,23ZaPeMc.exo}  is likely to be useful in helping to prioritise the molecules for which production of high-quality line lists suitable for the ExoMol database is warranted. Improved chemical models of a diverse set of atmospheres capable of predicting abundance of minor components with strong absorption would also be extremely valuable.    
With the planned launch of the Ariel space mission \cite{jt717,SPIE18} in 2029  producing a rapid increase in the number of exoplanet transit spectra available, we expect a corresponding increase in unidentified spectral features and thus the need to produce high-quality molecular spectral data on an increasing number of molecules \cite{jt946}. 

Yet, despite the launch of JWST, it is the ground-based observations that have generated the most substantial changes in the ExoMol database over the last four years as the high-resolution cross-correlation (HRCC) techniques has reached maturity. HRCC makes extremely high demands on the position accuracy of the strong spectral lines that are quite different from the high completeness required to model space observations. This difference has necessitated major changes in how line lists are created that had started in the 2020 ExoMol release with the introduction of MARVELized line lists but have become dominant in this 2024 release with the essential inclusion of energy level uncertainties within all line lists, major changes to line list construction methodology particularly the formalisation of the predicted shift and hybridisation approach, creation of the ExoMolHR web app, and even substantial updates to nuclear motion codes to enable modelling of hyperfine interactions. 

ExoMol was orginally designed to provide spectroscopic data at infrared and visible wavelengths.
However, increasing demands to study the effects of ultraviolet radiation both in terms
of photoabsorption and photodissociation has led to the scope of the database being extended
to address these issues. Processes involving temperature-dependent photodissociation,
line and continuum photoabsortion at UV wavelengths and predissociation are now being
included in the database. These data are needed both to interpret observations and for
chemical models of exoplanets. We are working on increasing our offering at UV wavelengths.


\section*{Data availability}

The data discussed in this paper can be accessed at www.exomol.com.
Programs associated with the ExoMol project including \textsc{ExoCross}  and \textsc{PyExoCross} are
available from https://github.com/ExoMol/.

\section*{Acknowledgments}
This work was  supported by the European Research Council (ERC) under the European Union's Horizon 2020 research and innovation programme through Advance Grant numbers 267219 and 883830, and   STFC Projects No. ST/I001050/1, ST/M001334/1 and ST/R000476/1, State Project IAP RAS No. FFUF-2024-0016


\bibliographystyle{elsarticle-num}

\appendix

\section{Sample JSON files}

\begin{python}[caption={JSON format master file \texttt{exomol.json}.}, label={lst:allJSON}]
{
    %%"ID"%%: "EXOMOL.master",
    %%"version"%%: 20240603,
    %%"num_molecules"%%: 91,
    %%"num_isotopologues"%%: 224,
    %%"num_datasets"%%: 106,
    %%"num_species"%%: 269,
    %%"molecules"%%: {
        %%"H2O"%%: {
            %%"num_molecule_names"%%: 11,
            %%"molecule_names"%%: [
                "Water",
                "Oxidane",
                "Hydrogen oxide",
                "Dihydrogen monoxide",
                "Hydrogen monoxide",
                "Dihydrogen oxide",
                "Hydrogen hydroxide",
                "Hydric acid",
                "Hydrohydroxic acid",
                "Hydroxic acid",
                "Hydrol"
            ],
            %%"num_isotopologues"%%: 7,
            %%"linelist"%%: [
                {
                    "inchikey": "XLYOFNOQVPJJNP-UHFFFAOYSA-N",
                    "iso_slug": "1H2-16O",
                    "iso_formula": "(1H)2(16O)",
                    "dataset_name": "POKAZATEL",
                    "version": 20180501
                },
                {
                    "inchikey": "XLYOFNOQVPJJNP-OUBTZVSYSA-N",
                    "iso_slug": "1H2-17O",
                    "iso_formula": "(1H)2(17O)",
                    "dataset_name": "HotWat78",
                    "version": 20161222
                },
                {
                    "inchikey": "XLYOFNOQVPJJNP-NJFSPNSNSA-N",
                    "iso_slug": "1H2-18O",
                    "iso_formula": "(1H)2(18O)",
                    "dataset_name": "HotWat78",
                    "version": 20161222
                },
                {
                    "inchikey": "XLYOFNOQVPJJNP-DYCDLGHISA-N",
                    "iso_slug": "1H-2H-16O",
                    "iso_formula": "(1H)(2H)(16O)",
                    "dataset_name": "VTT",
                    "version": 20160726
                },
                {
                    "inchikey": "XLYOFNOQVPJJNP-DYCDLGHISA-N",
                    "iso_slug": "1H-2H-16O",
                    "iso_formula": "(1H)(2H)(16O)",
                    "dataset_name": "Hewitt",
                    "version": 20161222
                },
                {
                    "inchikey": "XLYOFNOQVPJJNP-DYCDLGHISA-N",
                    "iso_slug": "1H-2H-16O",
                    "iso_formula": "(1H)(2H)(16O)",
                    "dataset_name": "TDB",
                    "version": 20240603
                },
                {
                    "inchikey": "XLYOFNOQVPJJNP-ZSJDYOACSA-N",
                    "iso_slug": "2H2-16O",
                    "iso_formula": "(2H)2(16O)",
                    "dataset_name": "Hewitt",
                    "version": 20161222
                }
            ]
        }
    }
}
\end{python}

\begin{python}[caption={JSON format definition (\texttt{.json}) file of ${^1}$H$_2^{16}$O POKAZATEL dataset.}, label={lst:H2OJSON}]
{
    %%"isotopologue"%%: {
        "iso_formula": "(1H)2(16O)",
        "iso_slug": "1H2-16O",
        "inchikey": "XLYOFNOQVPJJNP-UHFFFAOYSA-N",
        %%"inchi"%%: "1S/H2O/h1H2",
        %%"mass_in_Da"%%: 18.010565,
        %%"point_group"%%: "C2v"
    },
    %%"atoms"%%: {
        %%"number_of_atoms"%%: 3,
        %%"element"%%: {
            %%"H"%%: 1,
            %%"O"%%: 16
        }
    },
    %%"nuclear_spin_degeneracy"%%: {
        %%"A1"%%: 1,
        %%"A2"%%: 1,
        %%"B1"%%: 3,
        %%"B2"%%: 3
    },
    %%"dataset"%%: {
        "name": "POKAZATEL",
        "version": 20230621,
        %%"doi"%%: "10.1093/mnras/sty1877",
        %%"max_temperature"%%: 6000.0,
        %%"num_pressure_broadeners"%%: 4,
        %%"nxsec_files"%%: 0,
        %%"nkcoeff_files"%%: 0,
        %%"dipole_available"%%: false,
        %%"cooling_function_available"%%: false,
        %%"specific_heat_available"%%: true,
        %%"continuum"%%: null,
        %%"predis"%%: null,
        %%"states"%%: {
            %%"number_of_states"%%: 810269,
            %%"max_energy"%%: 41200.00,
            %%"uncertainties_available"%%: true,
            %%"lifetime_available"%%: false,
            %%"lande_g_available"%%: false,
            %%"num_quanta"%%: 6,
            %%"states_file_fields"%%: [
                {
                    "name": "ID",
                    "desc": "Unique integer identifier for the energy level",
                    "ffmt": "I12",
                    "cfmt": "%12d"
                },
                {
                    "name": "E",
                    "desc": "State energy in cm-1",
                    "ffmt": "F12.6",
                    "cfmt": "%12.6f"
                },
                {
                    "name": "gtot",
                    "desc": "Total energy level degeneracy",
                    "ffmt": "I6",
                    "cfmt": "%6d"
                },
                {
                    "name": "J",
                    "desc": "Total rotational quantum number, excluding nuclear spin",
                    "ffmt": "I7",
                    "cfmt": "%7d"
                },
                {
                    "name": "unc",
                    "desc": "Energy uncertainty in cm-1",
                    "ffmt": "F12.6",
                    "cfmt": "%12.6f"
                },
                {
                    "name": "nltcs:Ka",
                    "ffmt": "I2",
                    "cfmt": "%2d",
                    "desc": "Ka rotational quantum number"
                },
                {
                    "name": "nltcs:Kc",
                    "ffmt": "I2",
                    "cfmt": "%2d",
                    "desc": "Kc rotational quantum number"
                },
                {
                    "name": "nltcs:v1",
                    "ffmt": "I2",
                    "cfmt": "%2d",
                    "desc": "v1 symmetric stretch quantum number(3)"
                },
                {
                    "name": "nltcs:v2",
                    "ffmt": "I2",
                    "cfmt": "%2d",
                    "desc": "v2 bend quantum number"
                },
                {
                    "name": "nltcs:v3",
                    "ffmt": "I2",
                    "cfmt": "%2d",
                    "desc": "v3 asymmetrics stretch quantum number"
                },
                {
                    "name": "nltcs:Grve",
                    "ffmt": "A3",
                    "cfmt": "%3s",
                    "desc": "Rovibrational symmetry label"
                },
                {
                    "name": "Auxiliary:Ecal",
                    "desc": "Energy in cm-1 from variational spectroscopic model",
                    "ffmt": "F12.6",
                    "cfmt": "%12.6f"
                },
                {
                    "name": "Auxiliary:SourceType",
                    "ffmt": "A2",
                    "cfmt": "%2s",
                    "desc": "Indicates if the value is from MARVEL (Ma) or calculated (Ca)"
                }            
            ]
        },
        %%"transitions"%%: {
            %%"number_of_transitions"%%: 5745071340,
            %%"number_of_transition_files"%%: 412,
            %%"max_wavenumber"%%: 41200.00
        }
    },
    %%"partition_function"%%: {
        %%"max_partition_function_temperature"%%: 10000.0,
        %%"partition_function_step_size"%%: 1.0
    },
    %%"broad"%%: {
        %%"default_Lorentzian_half-width"%%: 0.07,
        %%"default_temperature_exponent"%%: 0.5,
        %%"H2"%%: {
            %%"filename"%%: "1H2-16O__H2.broad",
            %%"max_J"%%: 50,
            %%"Lorentzian_half_width"%%: 0.0209,
            %%"temperature_exponent"%%: 0.027,
            %%"quantum_number_sets"%%: [
                {
                    "code": "a3",
                    "num_lines": 23731,
                    "quantum_numbers": [
                        "J'",
                        "Ka\"",
                        "Ka'"
                    ]
                },
                {
                    "code": "a1",
                    "num_lines": 100,
                    "quantum_numbers": [
                        "J'"
                    ]
                },
                {
                    "code": "a0",
                    "num_lines": 51,
                    "quantum_numbers": []
                }
            ]
        },
        %%"He"%%: {
            %%"filename"%%: "1H2-16O__He.broad",
            %%"max_J"%%: 50,
            %%"Lorentzian_half_width"%%: 0.0042,
            %%"temperature_exponent"%%: 0.02,
            %%"quantum_number_sets"%%: [
                {
                    "code": "b1",
                    "num_lines": 253,
                    "quantum_numbers": [
                        "J'",
                        "Ka\"",
                        "Ka'",
                        "Kc\"",
                        "Kc'",
                        "v1\"",
                        "v2\"",
                        "v3\"",
                        "v1'",
                        "v2'",
                        "v3'"
                    ]
                },
                {
                    "code": "a3",
                    "num_lines": 23731,
                    "quantum_numbers": [
                        "J'",
                        "Ka\"",
                        "Ka'"
                    ]
                },
                {
                    "code": "a1",
                    "num_lines": 100,
                    "quantum_numbers": [
                        "J'"
                    ]
                },
                {
                    "code": "a0",
                    "num_lines": 51,
                    "quantum_numbers": []
                }
            ]
        }
    }
}
\end{python}

\end{document}